\documentclass[aps,pra,showpacs,reprint,amsmath,amssymb,onecolumn,notitlepage]{revtex4-1}

\usepackage{siunitx}
\usepackage{graphicx}
\usepackage{dcolumn}
\usepackage{bm}
\usepackage{hyperref}
\usepackage{epstopdf}
\usepackage{placeins}
\usepackage{amssymb}
\usepackage{color}
\usepackage{mathtools}

\begin{document}

\title{Frequency tuning of a triply-resonant whispering-gallery mode resonator to MHz wide transitions for proposed quantum repeater schemes}
\pdfoutput=1
\author{Gerhard Schunk$^{1,2,3,*}$}
\author{Ulrich Vogl$^{1,2}$}
\author{Florian Sedlmeir$^{1,2,3}$}
\author{Dmitry V. Strekalov$^{1,2}$}
\author{Alexander Otterpohl$^{1,2}$}
\author{Valentin Averchenko$^{1,2,3}$}
\author{Harald G. L. Schwefel$^{1,2,4}$}
\author{Gerd Leuchs$^{1,2}$}
\author{Christoph Marquardt$^{1,2,6}$}
\affiliation{$^{1}$Max Planck Institute for the Science of Light, G\"{u}nther-Scharowsky-Stra\ss e 1/Building 24, 90158 Erlangen, Germany}              
\affiliation{$^{2}$Institute for Optics, Information and Photonics, University Erlangen-N\"{u}rnberg, Staudtstr.7/B2, 90158 Erlangen, Germany}
\affiliation{$^{3}$SAOT, School in Advanced Optical Technologies, Paul-Gordan-Str. 6, 91052 Erlangen, Germany}
\affiliation{$^{4}$Department of Physics, University of Otago, 730 Cumberland Street, Dunedin 9016, New Zealand}
\affiliation{$^{5}$Department of Physics, Technical University of Denmark, Fysikvej Building 309, 2800 Lyngby, Denmark}
\affiliation{$^{*}$Corresponding author: Gerhard.Schunk@mpl.mpg.de}


\begin{abstract}
Quantum repeaters rely on interfacing flying qubits with quantum memories. The most common implementations include a narrowband single photon matched in bandwidth and central frequency to an atomic system. Previously, we demonstrated the compatibility of our versatile source of heralded single photons, which is based on parametric down-conversion in a triply-resonant whispering-gallery mode resonator, with alkaline transitions [Schunk \textit{et al.}, Optica \textbf{2}, 773 (2015)]. In this paper, we analyze our source in terms of phase matching, available wavelength-tuning mechanisms, and applications to narrow-band atomic systems. We resonantly address the D1 transitions of cesium and rubidium with this optical parametric oscillator pumped above its oscillation threshold. Below threshold, the efficient coupling of single photons to atomic transitions heralded by single telecom-band photons is demonstrated. Finally, we present an accurate analytical description of our observations. Providing the demonstrated flexibility in connecting various atomic transitions with telecom wavelengths, we show a promising approach to realize an essential building block for quantum repeaters.
\end{abstract}

\maketitle 


\tableofcontents

\section{Introduction}
Much progress has been achieved in the development of sources of quantum light. A source for which both central wavelength and bandwidth can be adapted to the optical transitions of e.g. single atoms \cite{Specht2011}, atomic ensembles \cite{Hald1999,Hammerer2005}, optomechanical resonators \cite{Aspelmeyer2014}, or quantum dots will give new prospects for a plethora of fundamental studies. Efficient photon-atom coupling paves the way for atom-based quantum memories \cite{Lvovsky2009,Simon2010}, super-radiance of collectively excited atoms \cite{R.H.Dicke1953}, two-photon spectroscopy \cite{Jechow2013}, and efficient photon-atom interaction either in free space \cite{Fischer} or in an optical cavity \cite{Ourjoumtsev2011a}.

Single photons from trapped atoms \cite{Legero2004,Volz2006,Beugnon2006,Maunz2007,Maiwald2012}, single molecules \cite{Siyushev2014}, rare earth ions in a crystal \cite{Utikal2014}, semiconductor quantum dots \cite{Akopian2011}, four-wave mixing in a cloud of cold atoms \cite{Leong2015}, and cavity-assisted parametric down-conversion (PDC) \cite{Bao2008,Wolfgramm2011,Fekete2013,Lenhard2015,Brecht2015,Schunk2015a} were demonstrated as promising candidates for photon-atom interactions. PDC has the technological advantage to not require an evacuated environment or cryostatic temperatures. At the single photon level PDC allows for one pump photon to split up into one pair of signal and idler photons. Nearly single-photon functionality of such a source is achieved by heralding: a single-photon detection event in e.g. the idler mode results in a heralded single photon in the signal mode. In general, the heralding  removes the vacuum component from the bipartite state also for non-photon-number-resolving click detection.

Optical resonators can be used to control the phase matching conditions \cite{Dunn1999}, to increase the conversion efficiency, and to reduce the bandwidth of the generated photons. We implement such an optical parametric oscillator (OPO) as a triply-resonant whispering-gallery mode resonator (WGMR), which confines light via total internal reflection. The small mode volumes, high quality factors, and the wavelength independent nature of total internal reflection make WGMRs an excellent platform for nonlinear optics \citep{Savchenkov2004,Del'Haye2007,Furst2010natural,Furst2010,Lin2013}. Evanescent field coupling allows to tune the pair generation rate and bandwidth of the generated photon pairs \cite{Gorodetsky1999}. Lithium niobate is a suitable host material due to its high $\chi^{(2)}$ nonlinearity and the possibility to achieve natural phase matching \cite{Furst2010natural}. Around the absorption minimum at a wavelength of \SI{1750}{\nm}, the intrinsic losses of lithium niobate and hence the minimal photon bandwidth can be as small as \SI{d-4}{\per \cm} or \SI{0.2}{MHz}, respectively \cite{Leidinger2015a}. For our investigated parametric wavelengths (approximately \SIrange{750}{1600}{\nm}) a minimal bandwidth of a few $\SI{}{\MHz}$ was observed.

Previously, we have reported on a parametric WGMR source of bright squeezed light \cite{Furst2011} and narrowband photon pairs \cite{Michael2013} compatible to atomic transitions \cite{Schunk2015a}. In this paper, we discuss the technical details crucial for the implementation and further development of such a source and explain its underlying principles. In the first part of the paper, we present our experimental scheme followed by an analysis of stepwise and continuous frequency tuning. We achieve continuous tuning of the parametric frequencies on the MHz scale with a movable dielectric substrate. Continuous tuning is used to match the parametric wavelength to the D1 lines of rubidium and cesium \cite{Schunk2015a}. The shown stability of the signal frequency is  comparable to the linewidth of the atomic transition. In a below-threshold experiment, we observe photons generated from atomic fluorescence that are heralded by idler parametric photons \cite{Schunk2015a}, for which we present a full analytic description.

\section{Characterization of PDC in a triply-resonant WGMR}

\subsection{Phase matching}
In this section we describe parametric interaction between three resonant fields in a WGMR.

\begin{figure}[htb]
	\centering
	\includegraphics[scale=0.8]{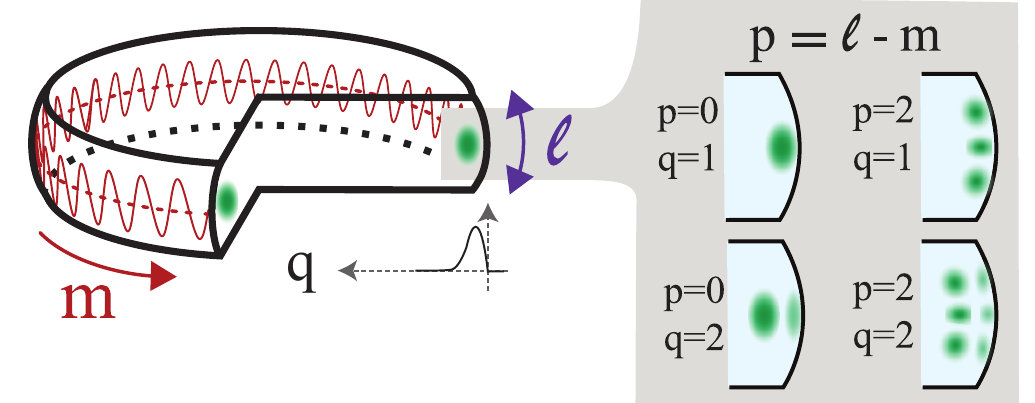}
	\caption{Illustration of the eigenmodes in a whispering-gallery mode resonator. These eigenmodes are defined by the azimuthal mode number m, the radial mode number q and the angular mode number p.}
\label{fig:modeillu}
\end{figure}

Solving Maxwell's equation for a dielectric spheroid, one obtains the eigenmodes of the WGMR. These eigenmodes are characterized by three integer numbers: the polar mode number $\ell \gg 1$, the azimuthal mode number $|\textrm{m}| \gg 1$, and the radial mode number $\textrm{q} \geqslant 1$. The angular mode number $\textrm{p}=\ell- |\textrm{m}|=0,1,2,...$ is additionally introduced to simplify the discussion of modes mostly used in the experiment (see Fig.~\ref{fig:modeillu}). The fundamental mode ($\textrm{q}_\textrm{}= 1, \textrm{p}_\textrm{}= 0$) shows in good approximation a Gaussian transversal profile with a single intensity maximum.

The eigenfrequencies of the system are defined by the dispersion relation \cite{Oraevsky2002,Gorodetsky2006}, which connects the mode numbers $\ell$, q, and $\textrm{p}$ with the resonance frequency $\nu(\ell,\textrm{q,p})$:
\begin{flalign}
 \nu({\ell,\textnormal{q,p}}) &= {\frac{c}{2\pi n R} } \cdot \left[ \ell + \alpha_\textnormal{q} \left( \frac{\ell}{2}\right)^{{1}/{3}} + p \left( \sqrt{\frac{R}{\rho}} - 1 \right) -\frac{\chi \cdot n}{\sqrt{n^2-1}} + \frac{1}{2}\sqrt{\frac{R}{\rho}} +  O\left( \ell^{-{1}/{3}} \right)  \right] \,, \label{eq:disprel}
\end{flalign}
where p and $\sqrt{R/\rho}$ are of the order of one. The WGMR radius and rim curvature are given by $R$ and $\rho$, respectively. The speed of light in vacuum is $c$. The parameter $\chi$ is 1 for TE modes and $1/n^2$ for TM modes, where the electric field is mainly orthogonal polarized for TE modes and mainly parallel polarized for TM modes with respect to the WGMR equatorial plane. The wavelength-dependent refractive index of the WGMR is $n$. The q-th root of the Airy function $\alpha_\textrm{q}>0$ can be approximated as $\alpha_\textrm{q} = [3 \pi/2  (\textrm{q} - 1/4) ]^{2/3}$. 

Phase matching of PDC in a triply-resonant WGMR requires energy conservation:
\begin{align}
	\nu_\textrm{p} &= \nu_\textrm{s} + \nu_\textrm{i} \,, \qquad 
	\label{eq:phasematcha}
\end{align}
expressed in terms of the oscillation frequencies $\nu_\textrm{p,s,i}$ of the electric fields, and momentum conservation \cite{Kozyreff2008,Furst2010natural}:
\begin{subequations}
	\begin{align}
	\textrm{m}_\textrm{p} &= \textrm{m}_\textrm{s} + \textrm{m}_\textrm{i} \,,\\
	\vert \textrm{m}_\textrm{s} + \textrm{p}_\textrm{s} - \textrm{m}_\textrm{i}  - \textrm{p}_\textrm{i} \vert &\leq \textrm{m}_\textrm{p} + \textrm{p}_\textrm{p}  \leq \textrm{m}_\textrm{s} + \textrm{p}_\textrm{s} + \textrm{m}_\textrm{i} + \textrm{p}_\textrm{i} \,,  \\
	 \textrm{m}_\textrm{p} + \textrm{m}_\textrm{s} + \textrm{m}_\textrm{i} &+ \textrm{p}_\textrm{p} + \textrm{p}_\textrm{s} + \textrm{p}_\textrm{i} \in 2\,\mathbb{Z}   \,,
	\label{eq:phasematchc}
	\end{align}
\label{eq:phasematch}
\end{subequations}
expressed in terms of the azimuthal mode numbers $\textrm{m}_\textrm{p,s,i}$ and the angular mode numbers $\textrm{p}_\textrm{p,s,i}$. The subscripts p, s, and i denote pump, signal, and idler, respectively. In case of $\nu_\textrm{p,s,i} = \nu({\ell_\textrm{p,s,i},\textnormal{q}_\textrm{p,s,i},\textrm{p}_\textrm{p,s,i}})$, the phase matching conditions given by Eq.~\ref{eq:phasematch} can be conveniently approximated by the waveguide phase matching condition:
\begin{align}
{n^{\prime}_\textrm{p} \nu_\textrm{p} = n^{\prime}_\textrm{s} \nu_\textrm{s} + n^{\prime}_\textrm{i} \nu_\textrm{i}} \,,
\label{eq:dispreleff}
\end{align} 
introducing the effective refractive index $n^\prime_\textrm{p,s,i} = {c \, \textrm{m}_\textrm{p,s,i}} / \left ({2 \pi R \, \nu_\textrm{p,s,i} }\right)$, which includes geometric dispersion to the refractive index. The azimuthal mode numbers $\textrm{m}_\textrm{p,s,i}$ follow from Eq.~\ref{eq:disprel}.

\begin{figure}[htb]
	\centering
	\includegraphics[scale=1.2]{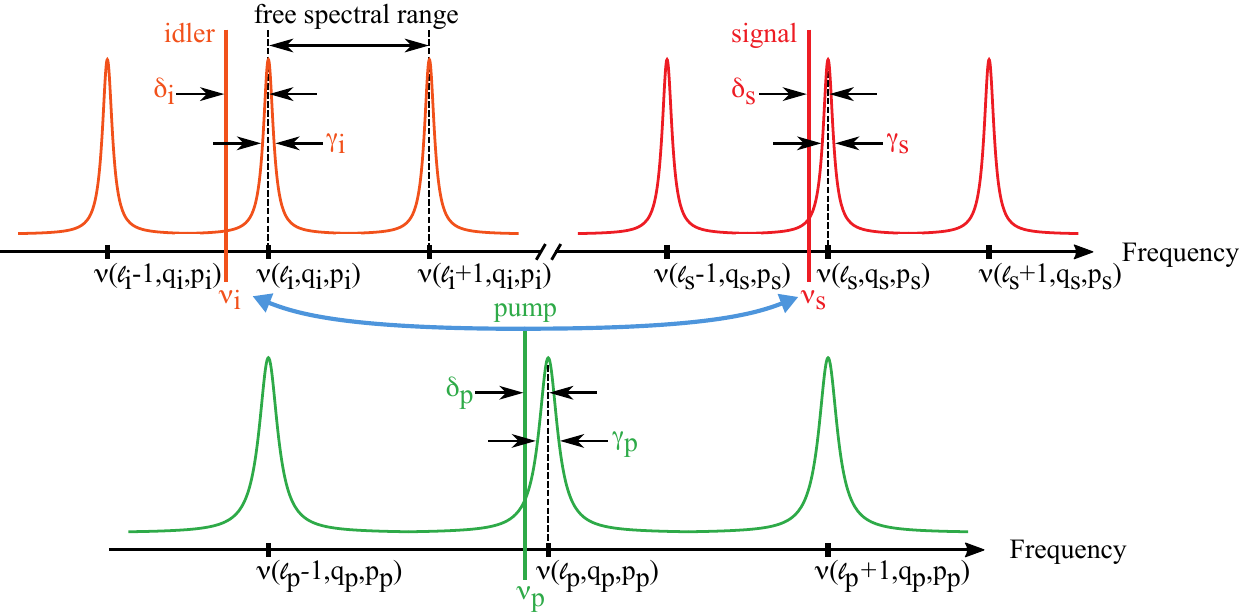}
	\caption{Frequency diagram of PDC in the triply-resonant case. In PDC the pump field with frequency $\nu_\textrm{p}$ is coupled to signal and idler fields with frequencies $\nu_\textrm{s}$ and $\nu_\textrm{i}$ under the condition of energy conservation given by Eq.~\ref{eq:phasematcha}. The respective frequency detuning $\delta_\textrm{p,s,i} = \left(\nu_\textrm{p,s,i} - \nu (\ell_\textrm{p,s,i},\textrm{q}_\textrm{p,s,i},\textrm{p}_\textrm{p,s,i})\right) / \gamma_\textrm{p,s,i}$ gives the normalized difference between the resonance frequency $\nu({\ell_\textrm{p,s,i},\textnormal{q}_\textrm{p,s,i},\textrm{p}_\textrm{p,s,i}})$ of a mode and the electric fields frequency $\nu_\textrm{p,s,i}$ exciting the modes. The respective bandwidths are given by $\gamma_\textrm{p,s,i}$.} 
\label{fig:SPDCmodesill}
\end{figure}

In general, the cavity resonances of signal and idler determine the frequency intervals in which electric fields can be excited by the PDC process (see Fig.~\ref{fig:SPDCmodesill}). The cavity resonances $g_\textrm{p,s,i}(\nu_\textrm{p,s,i})$ for pump, signal, and idler intensities are well approximated by Lorentzian functions:
\begin{align}
g_\textrm{p,s,i}(\nu_\textrm{p,s,i}) \propto \left[ { 1 + 4 \left({ \frac{\nu_\textrm{p,s,i} -\nu(\ell_\textrm{p,s,i},\textrm{q}_\textrm{p,s,i},\textrm{p}_\textrm{p,s,i})}{ \gamma_\textrm{p,s,i}} } \right)^2 } \right]^{-1} = \left[ {1+ 4  \delta_\textrm{p,s,i} ^2} \right]^{-1} \,.
\label{eq:cavityresponse}
\end{align} 
The respective frequency detuning $\delta_\textrm{p,s,i} = \left(\nu_\textrm{p,s,i} - \nu (\ell_\textrm{p,s,i},\textrm{q}_\textrm{p,s,i},\textrm{p}_\textrm{p,s,i})\right) / \gamma_\textrm{p,s,i}$ gives the normalized difference between the mode resonance frequency $\nu({\ell_\textrm{p,s,i},\textnormal{q}_\textrm{p,s,i},\textrm{p}_\textrm{p,s,i}})$ and the actual electric field frequency $\nu_\textrm{p,s,i}$ exciting the mode. The respective bandwidth $\gamma_\textrm{p,s,i} = \gamma^{\prime}_\textrm{p,s,i} + \gamma^{\prime\prime}_\textrm{p,s,i}$ is the sum of coupling rate $\gamma^{\prime}_\textrm{p,s,i}$ to the WGMR and loss rate $\gamma^{\prime\prime}_\textrm{p,s,i}$ in the WGMR. The coupling ratio $\kappa_\textrm{p,s,i} = {\gamma^{\prime}_\textrm{p,s,i}}/{\gamma^{}_\textrm{p,s,i}}<1$ gives the coupling scenario of the respective mode, i.e under-coupled for $\kappa_\textrm{p,s,i}<1/2$, critically coupled for $\kappa_\textrm{p,s,i}=1/2$, and over-coupled for $\kappa_\textrm{p,s,i}>1/2$.\\

The PDC frequency mismatch:
\begin{align}
 \Delta = \frac{\nu_\textrm{p} - \nu (\ell_\textrm{s},\textrm{q}_\textrm{s},\textrm{p}_\textrm{s} ) - \nu (\ell_\textrm{i},\textrm{q}_\textrm{i},\textrm{p}_\textrm{i} ) }{ \left( \gamma_\textrm{s}+\gamma_\textrm{i} \right)/2 } \,,
 \label{eq:SPDCfrequ}
\end{align} 
is the residual mismatch between the pump electric field frequency $\nu_\textrm{p}$ and the parametric resonance frequencies $\nu (\ell_\textrm{s,i},\textrm{q}_\textrm{s,i},\textrm{p}_\textrm{s,i})$ normalized to the mean bandwidth of signal and idler. In case of zero PDC mismatch, energy conservation (see Eq.~\ref{eq:phasematcha}) allows for a highly efficient conversion from the pump electric field frequency $\nu_\textrm{p}$ to the exact parametric resonance frequencies $\nu (\ell_\textrm{s,i},\textrm{q}_\textrm{s,i},\textrm{p}_\textrm{s,i})$. In a typical experiment, the pump laser frequency $\nu_\textrm{p}$ is locked to the resonance frequency of the pump mode ($\delta_\textrm{p}=0$) for a high intra-cavity power. The PDC frequency mismatch $\Delta$ is then controlled via the temperature-dependence of the resonance frequencies $\nu (\ell_\textrm{p,s,i},\textrm{q}_\textrm{p,s,i},\textrm{p}_\textrm{p,s,i})$.

The system exhibits a threshold with respect to the external pump power $P_\textrm{p}$: bright parametric beams are generated when the parametric gain exceeds the respective losses of signal and idler. This oscillation threshold $P_\textrm{th}\left( \delta_\textrm{p},\Delta\right)$ is given by \cite{Debuisschert1993,Richy1994,Breunig2013b}:
\begin{align}
	 P_\textrm{th} \left( \delta_\textrm{p},\Delta\right) = P_\textrm{0} \cdot \left(1 + 4 \delta_\textrm{p}^2 \right) \cdot \left(1 +  \Delta^2\right)   \,.
\label{eq:threshold}
\end{align}
In case of zero PDC mismatch ($\Delta=0$) and zero pump detuning ($\delta_\textrm{p}=0$), the minimal threshold $P_\textrm{0}$ is reached, which is typically of the order of a few \SI{}{\micro\watt} in our system \cite{Furst2010}. $P_\textrm{0}$ depends on the signal and the idler bandwidth, the coupling of the pump laser to the pump mode, the spatial overlap of pump, signal, and idler modes, and the $\chi^{(2)}$ nonlinearity coefficient, which we treat as constants in the following.

Far below the OPO threshold and in the limit of low gain, the rate of photon pair production in the resonator is given by:
\begin{align}
	{r_\textrm{si}}{} = 2 \pi \,  \frac{\gamma^{}_\textrm{s} \gamma^{}_\textrm{i}}{\gamma^{}_\textrm{s} + \gamma^{}_\textrm{i}}  \frac{P_\textrm{p}}{P_\textrm{th}\left( \delta_\textrm{p},\Delta \right)}  ,\quad (P_\textrm{p} \ll P_\textrm{th}) \,.
\label{eq:rategenbelow}
\end{align}
The coupling ratio $\kappa_\textrm{s,i}$ introduced above defines the escape probability for signal and idler photons from the resonator. The signal rate, the idler rate, and the pair generation rate outside the resonator are given by $R_\textrm{s} = \kappa_\textrm{s} \cdot r_\textrm{si}$, $R_\textrm{i} = \kappa_\textrm{i} \cdot r_\textrm{si}$, and $R_\textrm{si} =\kappa_\textrm{s} \cdot \kappa_\textrm{i} \cdot r_\textrm{si}$, respectively. 

Energy conservation does not result into monochromatic frequencies of signal and idler photons even in case of a monochromatic pump. In case of zero pump detuning, zero PDC mismatch, and equal bandwidths of signal and idler, the photons are generated over the full bandwidth of the respective modes \cite{Ou1999,predojevic2015engineering}.

Above the OPO threshold and in case of a monochromatic pump, bright signal and idler beams are generated at monochromatic frequencies. The so-called oscillation condition sets a joint frequency detuning $\delta_\textrm{si} = \delta_\textrm{s} = \delta_\textrm{i} $ for signal and idler, which is connected to the PDC frequency mismatch as  $\Delta = 2 \delta_\textrm{s,i} $. The signal/idler output power ${P_\textrm{s,i}}$ is given by \cite{Debuisschert1993,Richy1994,Breunig2013b}:
\begin{align}
	{P_\textrm{s,i}} =  4\, \kappa_\textrm{p}  \kappa_\textrm{s,i} P_\textrm{0} \frac{\nu_\textrm{s,i}}{\nu_\textrm{p}} \left(\sqrt{ \frac{P_\textrm{p}}{P_\textrm{0}} - \left(\Delta + 2 \delta_\textrm{p}\right)^2} - 1 + 2 {\delta_\textrm{p}} \Delta \right) ,\quad (P_\textrm{p} \geq P_\textrm{th}) \,.
	\label{eq:conveffabove}
\end{align}
In an experiment, the pump laser has a finite linewidth, which is typically much narrower than the bandwidths of the mode triplet. In this case, the linewidth of the parametric beams is determined by the linewidth of the pump.

\FloatBarrier

\subsection{Experimental scheme}
In this section we describe our experimental scheme for generation of non-classical light with PDC in a triply-resonant WGMR.

\begin{figure}[htb]
	\centering
	\includegraphics[scale=1]{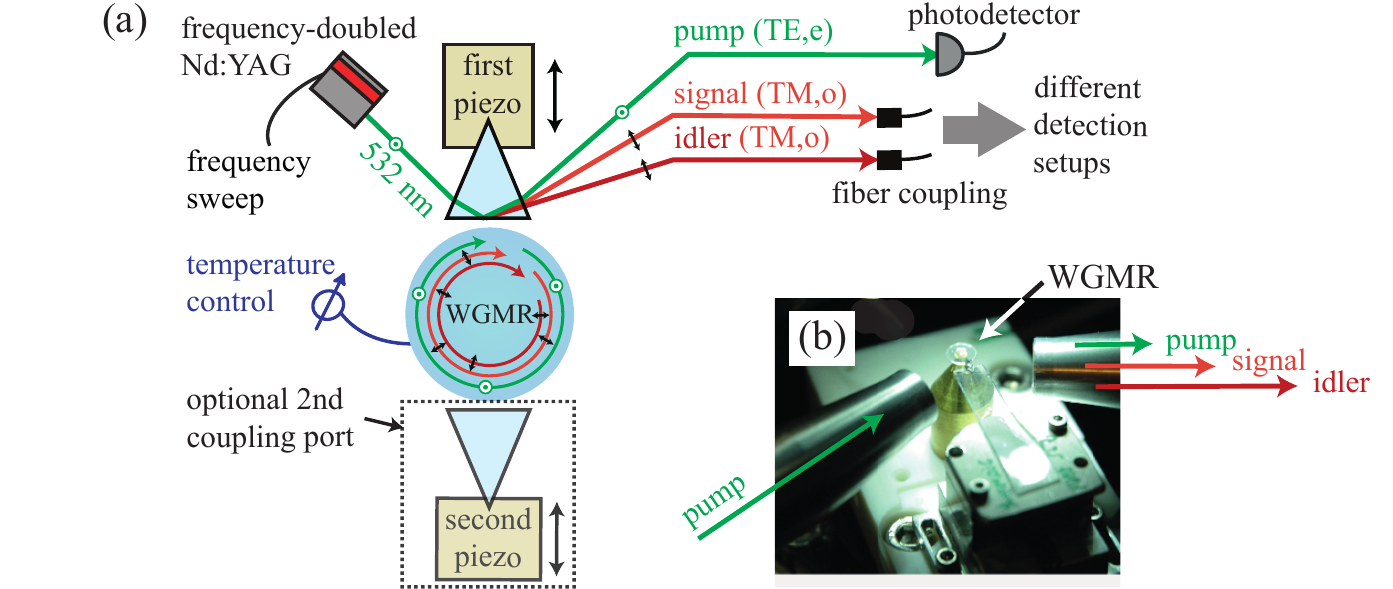}
	\caption{Experimental setup. (a) Squeezed light \cite{Furst2011} and photon pairs \cite{Michael2013} are generated via type-I parametric down-conversion in a triply-resonant whispering-gallery mode resonator. The pump beam couples to extraordinarily (e) polarized TE-modes. The parametric beams are generated in the ordinarily (o) polarized  TM-modes. A second coupling port is either used for an experimental identification of the modes with a second coupling prism \cite{Schunk2014a} or for tuning the resonance frequencies of the modes with a dielectric substrate \cite{Schunk2015a}. The generated pump, signal, and idler are separated with dichroic beam splitters. The pump is directly detected with a photodetector. The signal and idler are coupled to optical fibers and sent to different detection setups. (b) Image of the setup without temperature shielding and second coupling port.}
\label{fig:setup1}
\end{figure}

In the experiment (see Fig.~\ref{fig:setup1} and Ref.~\cite{Furst2010natural,Furst2010,Michael2013,Schunk2015a}) we use a WGMR made of MgO-doped (\SI{5.8}{\percent}) lithium niobate, which is manufactured by single point diamond turning and polishing techniques. For PDC, we pump the resonator with a frequency-doubled Nd:YAG laser (Prometheus, Innolight) at a wavelength of \SI{532}{\nm}, whose frequency detuning is measured by a Fabry-P\'{e}rot interferometer. The WGMR radius $R$ and rim curvature $\rho$ are \SI{2.5}{\mm} and \SI{0.58}{\mm}, respectively. We employ evanescent prism coupling \cite{Gorodetsky1999} to excite the pump mode and to couple the parametric beams to free space. The temperature of the WGMR is measured with a thermistor and stabilized with a proportional-integral-derivative controller at the millikelvin scale.

\begin{figure}[h]
	\centering
	\includegraphics[scale=1]{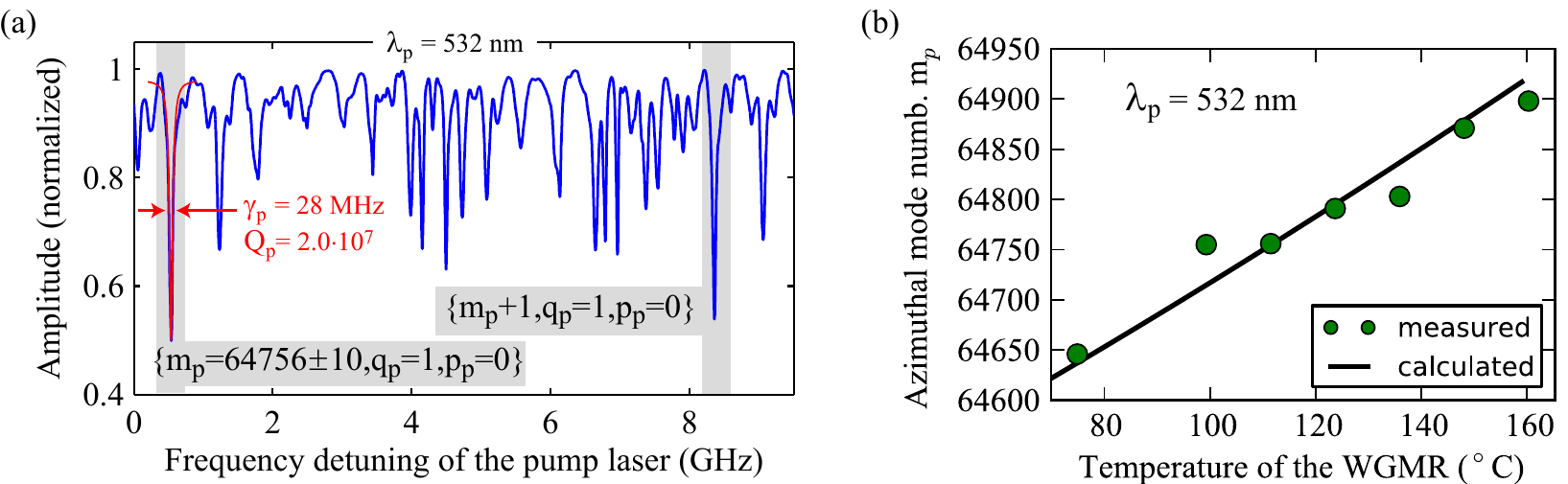}
	\caption{The temperature-dependent pump spectrum. (a) We measured the TE spectrum of the whispering-gallery mode resonator at the pump wavelength $\lambda_\textrm{p}=\SI{532}{\nm}$ over more than a free spectral range at a temperature of $T=\SI{100}{\degreeCelsius}$. A Lorentzian fit (red trace) of the fundamental mode yields the bandwidth of $\gamma_\textrm{p}=28\,\textrm{MHz}$ and the quality factor of $Q_\textrm{p}=2.0 \cdot 10^7 $ at critical coupling. ( (b) Temperature tuning of the WGMR changes the pump spectrum according to the thermal expansion \cite{Weis1985} and the thermo-optic \cite{Schlarb1994} coefficients of lithium niobate. At a given wavelength of the pump laser we changed the WGMR temperature by $\Delta T=\SI{70}{\degreeCelsius}$ and measured a change in the azimuthal mode number $\textrm{m}_\textrm{p}$ of the fundamental pump mode \cite{Schunk2014a} by approximately $\Delta\textrm{m}_\textrm{p}=250$.}
\label{fig:pumpspectrum}
\end{figure}

From the analysis of the spectral positions and far-field patterns of the pump modes,we identify the mode numbers as described in Ref.~\cite{Schunk2014a}. The frequency spectrum in Fig.~\ref{fig:pumpspectrum}(a) is measured at a WGMR temperature of \SI{100}{\degreeCelsius}. For this spectrum, the analysis yields an azimuthal mode number $\textrm{m}_\textrm{p}= 64756 \pm 10 $ for the fundamental pump mode. 

For higher WGMR temperatures, the product of WGMR radius and refractive index ${ Rn}$ increases since thermal refraction ${d n}/{d T}$ \cite{Schlarb1994} and thermal expansion \cite{Weis1985} ${d R}/{d T}$ are both positive functions in case of lithium niobate. Keeping the frequency sweep range of the laser fixed, this temperature induced change of refractive index leads to a coupling of the pump laser to modes with higher azimuthal mode numbers $\textrm{m}$ according to Eq.~\ref{eq:disprel}. In Fig.~\ref{fig:pumpspectrum}(b) the measured azimuthal mode number $\textrm{m}_\textrm{p}$ of a fundamental pump mode is shown at various temperatures together with the calculated values.

Phase matching in the WGMR depends on the refractive indices of the host material. In the degenerate case ($\nu_\textrm{p}=2 \cdot\, \nu_\textrm{s,i}$), the effective refractive indices $n^\prime_\textrm{p,s,i}$ of all three modes have to be equal, which follows from Eq.~\ref{eq:dispreleff}. To fulfill these conditions for three resonator modes, material dispersion must be compensated. We achieve this with natural phase matching of type-I PDC in a z-cut configuration of the lithium niobate crystal where the optic axis is aligned along the symmetry axis of the resonator \cite{Furst2010natural}. We pump the extraordinarily polarized TE modes of the WGMR, while the parametric light is generated in the ordinarily polarized TM modes. Due to the negative birefringence of lithium niobate, where the refractive index for the ordinary axis is higher than for the extraordinary axis, refractive index matching and therefore phase matching can be fulfilled. This concept extends to the case of non-degenerate PDC ($\nu_\textrm{i}< \nu_\textrm{s}$).

We use the fundamental mode for pumping the resonator. In principle, all eigenmodes with significant coupling (see Fig.~\ref{fig:pumpspectrum}(a)) can be used for PDC. However, a conversion from a fundamental pump to fundamental parametric modes exhibits the lowest minimal threshold $P_\textrm{0}$ given by Eq~\ref{eq:threshold} \cite{Furst2010}. In terms of the generated wavelength, this fundamental conversion channel is well separated from other conversion channels with higher radial or angular modes (see Sec.~\ref{sec:selmodetrip}). This separation is beneficial for single mode operation of the photon pair source \citep{Michael2014}.

\subsection{Selecting mode triplets by temperature}
\label{sec:selmodetrip}
In this section we discuss phase matching of PDC in WGMRs for the various mode combinations. Whispering gallery modes are guided almost entirely within the dielectric via total internal reflection. The wavelength-independent nature of total internal reflection allows pump, signal, and idler to be resonantly enhanced (triply-resonant OPO). This narrows down the phase matching conditions based on the small parametric bandwidths and the condition of momentum conservation. PDC in bulk lithium niobate is highly tunable by changing the crystal temperature. In the following, we explain how the discrete PDC spectrum in a WGMR can also be tuned via this temperature dependence.

\begin{figure}[htb]
	\centering
	\includegraphics[scale=1]{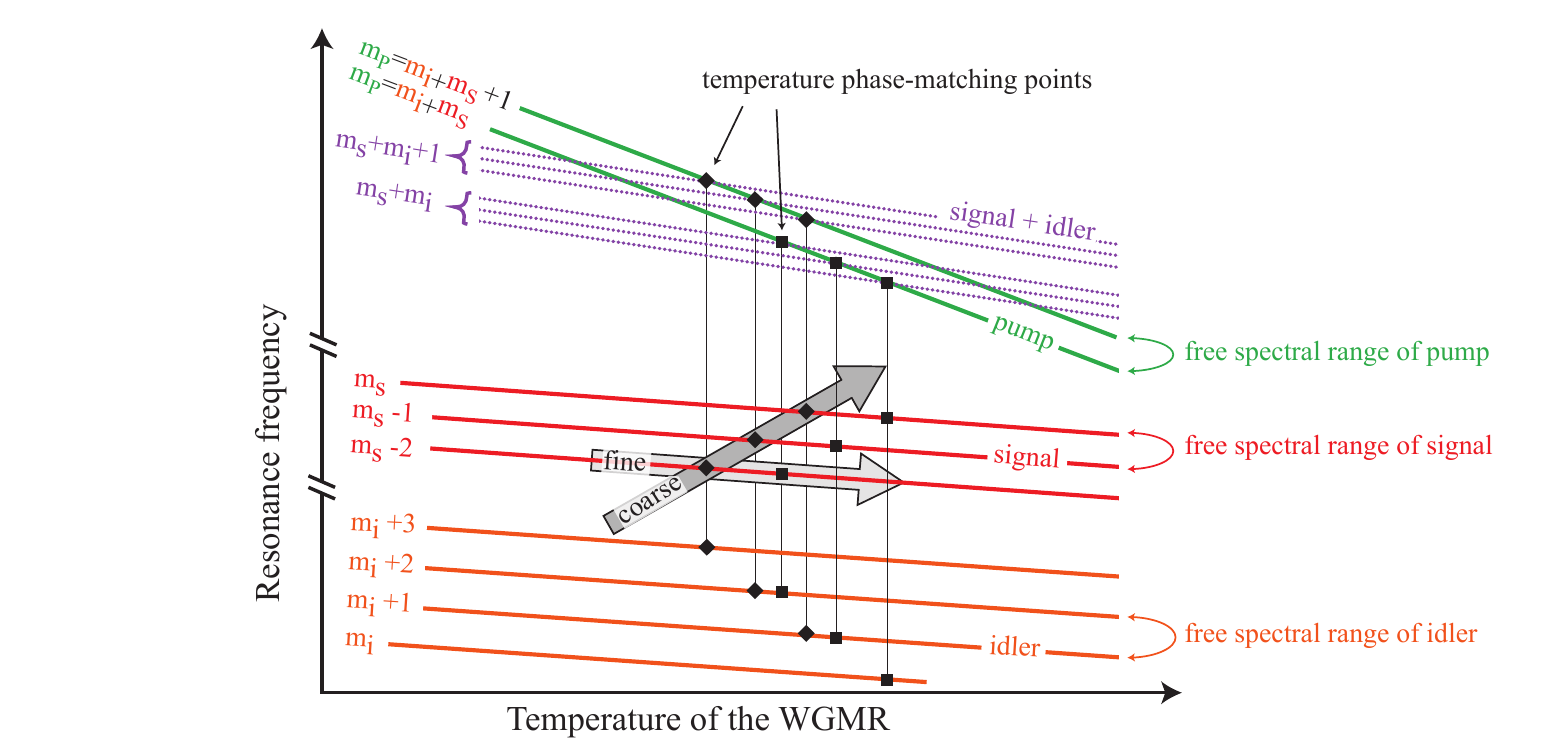}
	\caption{Selecting azimuthal parametric modes by temperature in a triply-resonant whispering-gallery mode resonator. Phase matching requires energy conservation given by Eq.~\ref{eq:phasematcha} and momentum conservation given by Eq.~\ref{eq:phasematch}. Energy conservation for a mode triplet is fulfilled at the intersection of the pump resonance frequency  $\nu(\ell_\textrm{p},\textrm{q}_\textrm{p},\textrm{p}_\textrm{p})$ (solid green) and the combined signal and idler resonance frequency $\nu(\ell_\textrm{s},\textrm{q}_\textrm{s},\textrm{p}_\textrm{s}) + \nu(\ell_\textrm{i},\textrm{q}_\textrm{i},\textrm{p}_\textrm{i})$ (dashed purple). The first condition of momentum conservation (Eq.~\ref{eq:phasematch}(a)) is fulfilled when the azimuthal mode numbers $\textrm{m}_\textrm{s}$ and $\textrm{m}_\textrm{i}$ of signal and idler, respectively, add up to the azimuthal mode number $\textrm{m}_\textrm{p}$ of the pump. Keeping the same pump (signal) mode while changing the idler and signal (pump) mode in our experiment leads to a tuning of the signal frequency in coarse (fine) steps indicated by the dark (light) grey arrow.} 
\label{fig:tempillstration}
\end{figure}

The basic principle behind natural phase matching in a triply-resonant WGMR is explained in Fig.~\ref{fig:tempillstration}. The temperature-dependent resonance frequencies $\nu(\ell_\textrm{p,s,i},\textrm{q}_\textrm{p,s,i},\textrm{p}_\textrm{p,s,i})$ of different azimuthal pump, signal, and idler modes are depicted as solid green, red, and orange lines, respectively. The radial mode numbers $\textrm{q}_\textrm{p,s,i}$ and angular mode number $\textrm{p}_\textrm{p,s,i}$ for pump, signal, and idler, respectively, are the same for different azimuthal modes. As an illustration of energy conservation given by Eq.~\ref{eq:phasematcha}, we show the sum of the signal and idler resonance frequencies $\nu(\ell_\textrm{s},\textrm{q}_\textrm{s},\textrm{p}_\textrm{s})+\nu(\ell_\textrm{i},\textrm{q}_\textrm{i},\textrm{p}_\textrm{i})$ as purple dashed lines. Energy conservation in the case of zero detuning ($\delta_\textrm{p,s,i}=0$, cf. Fig.~\ref{fig:SPDCmodesill}) is fulfilled at the intersection of any purple dashed line with any solid green line. Phase matching is only found when the azimuthal mode numbers of signal and idler add up to the azimuthal mode number of the pump, which is the first condition of momentum conservation given by Eq.~\ref{eq:phasematch}(a).

There are two methods to tune the azimuthal mode frequency by temperature. The first method is to select a specific pump mode and to track it with the pump laser while changing the resonator temperature. The azimuthal mode numbers  $\textrm{m}_\textrm{s}$ and $\textrm{m}_\textrm{i}$ of signal and idler, respectively, will change in unity steps ($\Delta \textrm{m}_\textrm{s} \pm 1,\Delta \textrm{m}_\textrm{i} \mp1$). The second method is to change the azimuthal pump mode's number by unity steps, $\textrm{m}_\textrm{p} \rightarrow \textrm{m}_\textrm{p} \pm 1$, and adjust the temperature to restore the phase matching for a selected (e.g., signal) mode $\textrm{m}_\textrm{s}$. This leads to unity steps of the idler mode, $\textrm{m}_\textrm{i} \rightarrow \textrm{m}_\textrm{i} \pm 1$, while $\textrm{m}_\textrm{s}$ remains unchanged.

In an experiment, the first method is more intuitive since it requires only slight changes in the pump laser frequency following the temperature-dependent frequency shift of the pump mode. The second method starts with a significant change in the pump laser frequency by one free spectral range. Subsequently, a searching of the initial signal azimuthal mode via the first tuning method is required. For this, the frequency of the generated signal can be used as an observable.

\begin{figure}[htb]
	\centering
	\includegraphics[scale=1]{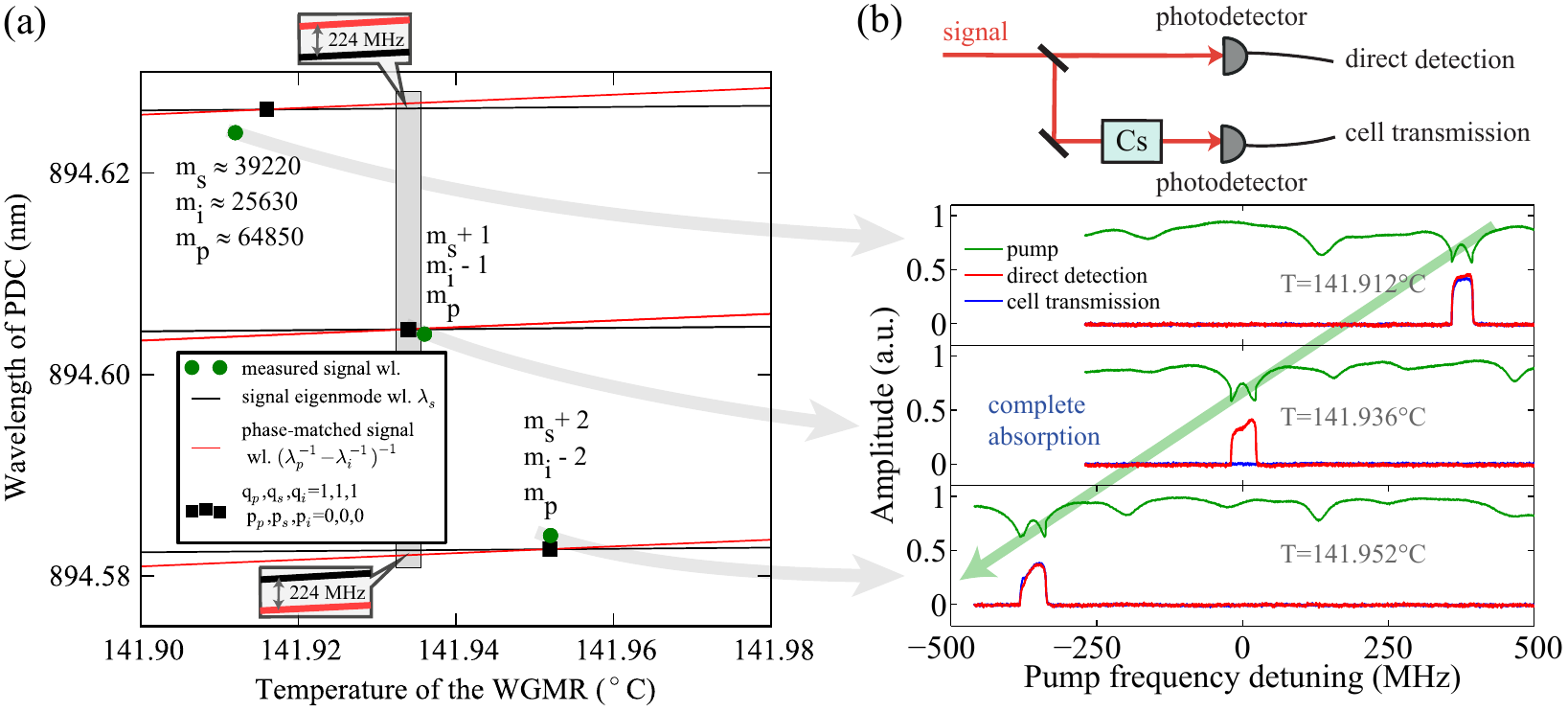}
	\caption{Selection of the parametric mode depending on the resonator temperature. (a) Energy conservation  (see Eq.~\ref{eq:phasematcha}) for one mode triplet requires that the resonance of the signal mode at $\lambda_\textrm{s}$ is equal to $(\lambda^{-1}_\textrm{p}-\lambda^{-1}_\textrm{i})^{-1}$. Phase matching can only be found for certain temperatures based on the nearly linear dependence on temperature of the resonance frequencies. (b) Coarse tuning observation setup and result. We split the generated signal equally into two beams, where one beam is sent through a Cs cell and the other detected directly. A sweep of the pump laser frequency at three different temperatures of the resonator is shown together with the generated signal for this triply-resonant OPO pumped highly above the threshold. The pump mode shifts continuously with temperature, but parametric light appears and disappears. Nearly full absorption of the generated signal for the resonator temperature \SI{141.936}{\degreeCelsius} indicates frequency-matching of Cs D1 transition, see Ref.~\cite{Schunk2015a}.} 
\label{fig:finetuning}
\end{figure}

For our experiment, the first method provides a coarse tuning (see Fig.~\ref{fig:finetuning}) for the signal and idler frequency in steps of $\pm\SI{8.2}{\GHz}$ and $\mp\SI{8.4}{\GHz}$, respectively. This corresponds approximately to the respective free spectral ranges. The second method enables fine tuning (see Fig.~3(a) in Ref.~\cite{Schunk2015a}) of either the signal frequency in steps of $\SI{254}{\MHz}$ with a constant $\textrm{m}_\textrm{s}$ or the idler frequency in steps of $\SI{130}{\MHz}$ with a constant $\textrm{m}_\textrm{i}$. In both cases these steps are considerably smaller than the free spectral ranges.

The temperature-dependent phase matching pertaining to our experiment is shown in Fig.~\ref{fig:finetuning}. In (a), the calculated resonance wavelength $\lambda_\textrm{s}$ of the fundamental signal mode is shown together with the signal wavelength  $(\lambda^{-1}_\textrm{p}-\lambda^{-1}_\textrm{i})^{-1}$ found from energy conservation given by Eq.~\ref{eq:phasematcha}. The resonance wavelengths of the pump and the idler are given by $\lambda_\textrm{p,i}$, respectively. PDC occurs for a triplet of modes fulfilling the momentum conservation conditions (see Eq.~\ref{eq:disprel}(a)-(c)), only when the phase-matched signal frequency intersects with an actual WGMR resonance frequency (cf. Fig.~\ref{fig:tempillstration}). 

A frequency sweep over the fundamental pump mode and the generated signal above threshold is shown in Fig.~\ref{fig:finetuning}(b) for three different temperatures. The resonance frequencies of the modes change continuously with temperature, whereas PDC is only generated for distinct temperature and pump frequency combinations (see Fig.~\ref{fig:tempillstration} and Fig.~\ref{fig:finetuning}). The non-Lorentzian pump resonance is well described by additional losses caused by the parametric process \cite{Breunig2013b}. As shown in Fig.~\ref{fig:finetuning}(a), individual steps originate from stepwise changes in the azimuthal mode number $\textrm{m}_\textrm{s,i}$ of signal and idler (cf. Fig.~\ref{fig:tempillstration}), keeping the pump laser frequency locked to the fundamental mode.

In doubly-resonant waveguides (cf. Fig.~2 in Ref.~\cite{Brecht2015}), a simultaneous excitation of various azimuthal modes of signal and idler separated by multiple free spectral ranges is still possible below threshold. In our triply-resonant case, this is prevented by the momentum conservation condition given by Eq.~\ref{eq:phasematch}. However, when one PDC channel is fully resonant (PDC frequency mismatch $\Delta =\SI{0}{}$), an unwanted generation of PDC to the adjacent azimuthal signal and idler modes is possible below the OPO threshold. The corresponding pair generation rates can be estimated for the example depicted in Fig.~\ref{fig:finetuning}(a). According to Eq.~\ref{eq:rategenbelow}, the pair generation rate to a signal and idler mode combination $\textrm{m}_\textrm{p},\textrm{m}_\textrm{s}+1,\textrm{m}_\textrm{i}-1$ with zero PDC frequency mismatch $\Delta =\SI{0}{}$ is 1153 times higher at our minimal photon bandwidth of $\gamma_\textrm{s}=\gamma_\textrm{i}=\SI{6.6}{\MHz}$ (see Fig.~\ref{fig:BWtuning}) than the pair generation rate to an adjacent azimuthal signal and idler mode combination with $\textrm{m}_\textrm{p},\textrm{m}_\textrm{s},\textrm{m}_\textrm{i}$ with the calculated mismatch $\nu (\ell_\textrm{p},\textrm{q}_\textrm{p},\textrm{p}_\textrm{p}) - \nu (\ell_\textrm{s},\textrm{q}_\textrm{s},\textrm{p}_\textrm{s}) - \nu (\ell_\textrm{i},\textrm{q}_\textrm{i},\textrm{p}_\textrm{i}) =\SI{224}{\MHz}$.

\begin{figure}[htb!]
	\centering
	\includegraphics[scale=1]{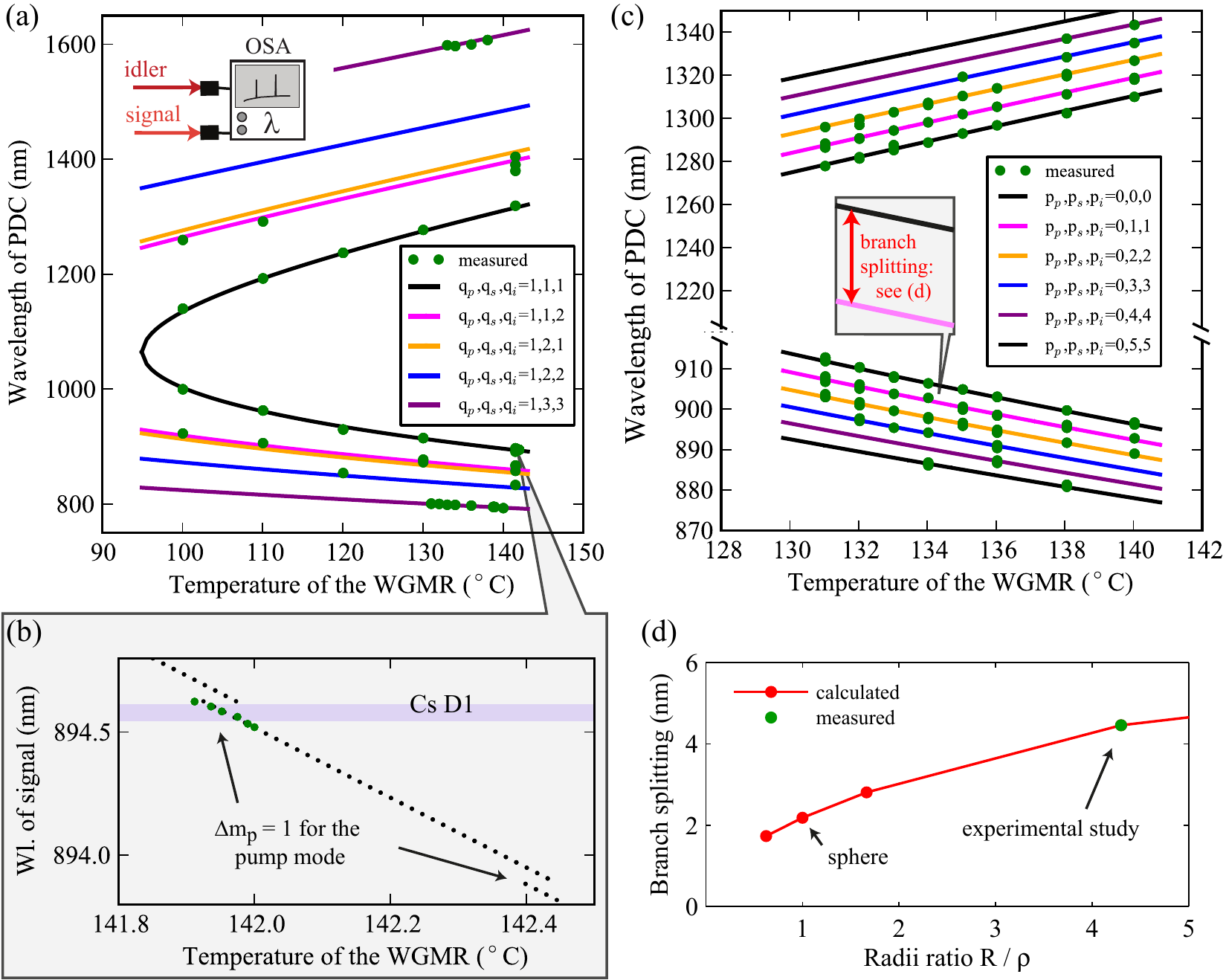}
	\caption{Measured and calculated temperature tuning of the parametric wavelengths using the fundamental pump mode with $\textrm{m}_\textrm{p} \gg 1$, $\textrm{q}_\textrm{p}=1$, and $\textrm{p}_\textrm{p}=0$. (a) The experimental results are shown together with calculations for varying equatorial mode numbers $\textrm{q}_\textrm{s,i}$ and equal polar mode numbers $\textrm{p}_\textrm{s,i}=0$ for signal and idler. Resonant light for the cesium D1 and rubidium D1 transition are generated in the $\textrm{q}_\textrm{p},\textrm{q}_\textrm{s},\textrm{q}_\textrm{i}=1,1,1$ and the $\textrm{q}_\textrm{p},\textrm{q}_\textrm{s},\textrm{q}_\textrm{i}=1,3,3$ channels, respectively (see Fig.~\ref{fig:beatsignals} and Ref.~\cite{Schunk2015a}). (b) The stepwise tuning behavior of the signal wavelength is due to a stepwise change in the azimuthal mode number $\textrm{m}_\textrm{s,i}$ of signal and idler (see Fig.~\ref{fig:tempillstration} and Fig.~\ref{fig:finetuning}(b)). (c) The measurement and calculations for varying polar mode numbers $\textrm{p}_\textrm{s,i}$ and equal radial mode numbers $\textrm{q}_\textrm{s,i}=1$ for signal and idler show a nearly equidistant splitting between the different branches. Standard interference filters can be used to filter out the fundamental conversion channel for single mode operation \cite{Michael2014}. (d) The splitting between the fundamental conversion channel ($\textrm{p}_\textrm{s} = 0, \textrm{p}_\textrm{i} = 0$) and the next higher branch ($\textrm{p}_\textrm{s} = 1, \textrm{p}_\textrm{i} = 1$) strongly depends on the ratio $R/\rho$ of the WGMR radius $R$ and curvature $\rho$ ($R/\rho\,=\,4.3$ in our experimental study).}
\label{fig:convchannels}
\end{figure}

\begin{figure}[htb]
	\centering
	\includegraphics[scale = 1]{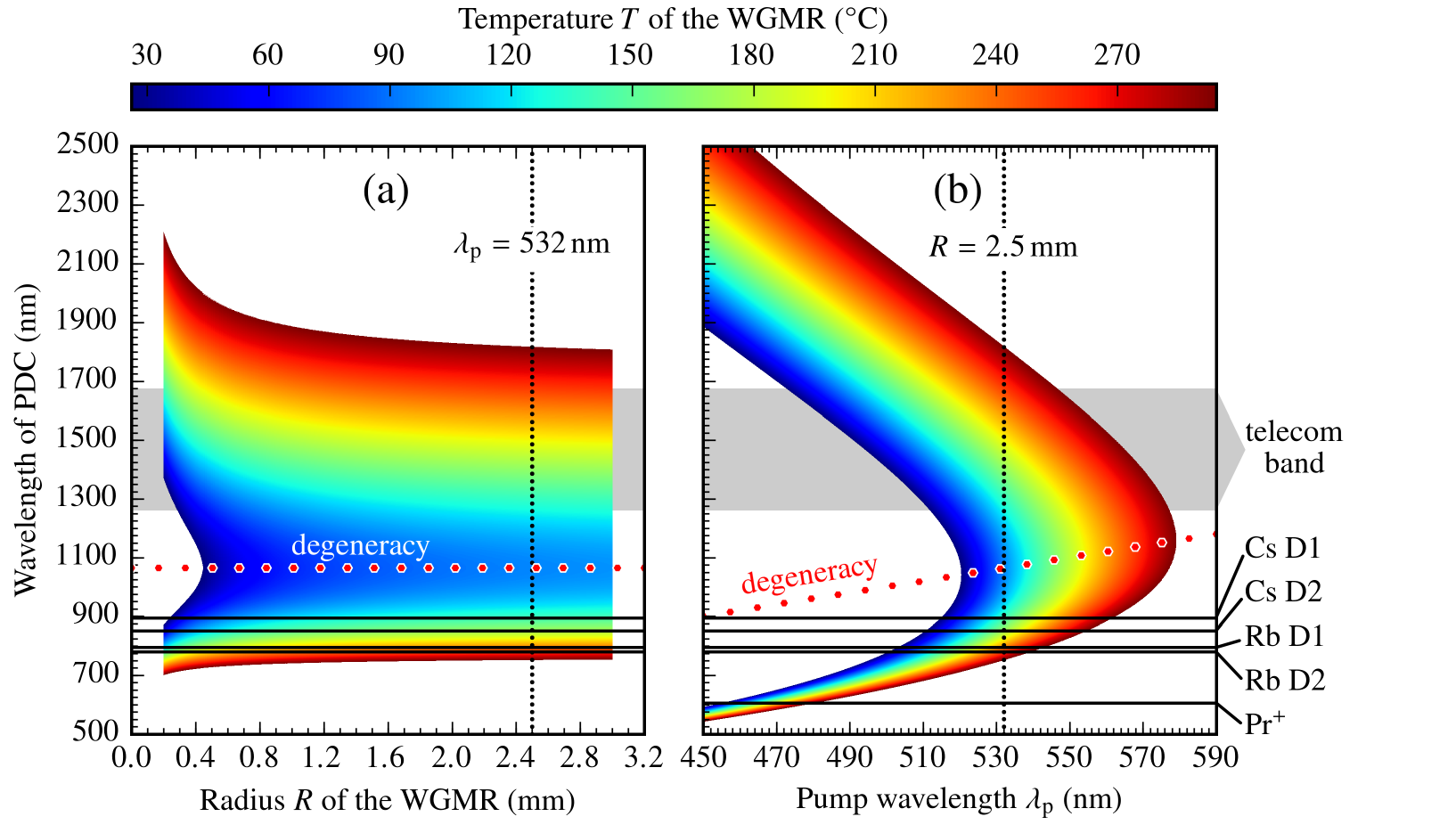}
	\caption{Calculations of the phase-matched wavelengths in parametric down-conversion for different resonator radii R at a pump wavelength $\lambda_\textrm{p}=\SI{532}{\nm}$ in (a) and for different pump wavelengths $\lambda_\textrm{p}$ for a resonator with radius R=\SI{2.5}{\mm} in (b). Fundamental pump, signal, and idler modes with  $\textrm{m}_\textrm{p,s,i} \gg 1$, $\textrm{q}_\textrm{p,s,i}=1$, and $\textrm{p}_\textrm{p,s,i}=0$ were used for both calculations (see Fig.~\ref{fig:convchannels} for higher-order conversion channels). Waveguide dispersion in (a) is more significant for smaller resonators, which can lead to a strong decrease of the phase matching temperatures. Larger resonators, however, simplify a tuning to the different narrowband transitions due to the smaller free spectral range. The strong dependence of the PDC process on the pump wavelength in (b) mainly arises from material dispersion in lithium niobate \cite{Schlarb1994}. The telecom band in the infrared for low-loss fiber communication and resonant wavelengths of cesium and rubidium atoms \cite{Steck2008all}, and praseodymium ions \cite{Utikal2014} are shown as horizontal lines.}
\label{fig:tempsizes}
\end{figure}

All temperature-dependent calculations for our system (Fig.~\ref{fig:pumpspectrum}(b), Fig.~\ref{fig:finetuning}, Fig.~\ref{fig:convchannels}, and Fig.~\ref{fig:tempsizes}) are based on thermorefractivity \cite{Schlarb1994} and thermal expansion \cite{Weis1985} of lithium niobate. Limited knowledge on the actual MgO concentration of the crystal, the actual temperature at the rim of the resonator, and the uncertainty on temperature induced strain on the resonator, which is glued to a base plate, make it necessary to introduce an additional scaling for the calculations. Therefore, we converted the calculated phase matching temperatures as ${T \rightarrow 1.22 \cdot T  + \SI{11}{\degreeCelsius}}$ to match the calculations to the measurement results in the investigated temperature range from \SIrange{70}{160}{\degreeCelsius}, which then agree consistently with the experimental data.

Phase matching depends on many different parameters such as the WGMR temperature, the pump wavelength, the respective mode triplet, the resonator size, and the MgO concentration of lithium niobate, or more generally, on the WGMR host material properties. For planning an experiment, an efficient numerical characterization of the parameter space is required. We have developed an algorithm to examine phase matching for a selected combination of pump, signal, and idler modes with given radial and angular mode numbers $\textrm{q}_\textrm{p,s,i}$ and $\textrm{p}_\textrm{p,s,i}$, respectively. The WGMR radius $R$, the rim curvature $\rho$, the pump wavelength range (approximately one free spectral range, see Fig.~\ref{fig:pumpspectrum}(a)), and the investigated temperature range are preset. The algorithm is designed to find the exact phase-matched resonance frequencies $\nu(\ell_\textrm{p,s,i},\textrm{q}_\textrm{p,s,i},\textrm{p}_\textrm{p,s,i})$ (zero detuning, $\delta_\textrm{p,s,i}=0$) given by Eq.~\ref{eq:disprel} and WGMR temperature for each combination of the azimuthal mode numbers $\textrm{m}_\textrm{p,s,i}$, constrained by the energy conservation condition given by  Eq.~\ref{eq:phasematcha} and momentum conservation given by Eq.~\ref{eq:phasematch}. 

The algorithm works as follows. First, the approximate phase-matched parametric frequencies are calculated by fixing the WGMR temperature and pump wavelength and treating the azimuthal mode numbers $\textrm{m}_\textrm{p,s,i}$ as continuous variables. The algorithm determines the azimuthal mode number $\textrm{m}_\textrm{p}$ of the pump and then varies the azimuthal mode numbers $\textrm{m}_\textrm{s}$ of the signal until energy conservation is fulfilled. The corresponding azimuthal mode numbers $\textrm{m}_\textrm{i}$ of the idler follow from momentum conservation. In a next step the azimuthal mode numbers $\textrm{m}_\textrm{p,s,i}$ are rounded to integer numbers while obeying momentum conservation, which results in shifted resonance frequencies $\nu(\ell_\textrm{p,s,i},\textrm{q}_\textrm{p,s,i},\textrm{p}_\textrm{p,s,i})$. Finally, the resulting energy mismatch is compensated by varying the WGMR temperature to yield the exact phase matching temperatures and resonance frequencies $\nu(\ell_\textrm{p,s,i},\textrm{q}_\textrm{p,s,i},\textrm{p}_\textrm{p,s,i})$ of one phase-matched triplet. 

Various PDC conversion channels originating from a fundamental pump mode are presented in Fig.~\ref{fig:convchannels} for the temperature ranging from \SI{100}{\degreeCelsius} to \SI{140}{\degreeCelsius}. Geometric dispersion leads to a drastic change in the emission frequency for the signal and the idler modes with different radial mode numbers $\textrm{q}_\textrm{s,i}$ in (a) and different angular mode numbers $\textrm{p}_\textrm{s,i}$ in (c). Each conversion channel in (c) with $\textrm{p}_\textrm{s}=\textrm{p}_\textrm{i}$ represents a nearly degenerate cluster of conversion channels generally defined by the cluster number $a = (\textrm{p}_\textrm{s}+\textrm{p}_\textrm{i} - \textrm{p}_\textrm{p} )\in 2\,\mathbb{N}_0$ \cite{Michael2014}, which are not shown for simplicity.

While changing the temperature, we keep the frequency of the pump within one free spectral range of the resonator (see Fig.~\ref{fig:pumpspectrum}(a)) due to the constraints on frequency tuning of the pump laser. Hence, the azimuthal mode number $\textrm{m}_\textrm{p}$ of the pump mode has to be changed by one after a certain temperature interval. This results in the two discontinuities shown in Fig.~\ref{fig:convchannels}(b) (cf. Fig.~\ref{fig:pumpspectrum}(b)). The conversion to modes with higher radial mode numbers $\textrm{q}_\textrm{s,i}$ and angular mode numbers $\textrm{p}_\textrm{s,i}$ needs to be suppressed or filtered out for single-mode operation \cite{Michael2014}, which is typically done using standard interference filters with a bandpass window of a few nanometer. The wavelength difference of the fundamental conversion channel $\textrm{p}_\textrm{s,i} = 0$ to higher order angular conversion channels $\textrm{p}_\textrm{s,i} \neq 0$ strongly depends on the curvature $\rho$ of the WGMR (see Fig.~\ref{fig:convchannels}(d)). 

For tuning the parametric wavelengths to various other physical systems we analyze the most efficient conversion channel from a fundamental pump to fundamental signal and idler modes in Fig.~\ref{fig:tempsizes} in terms of the radius R of the WGMR, the central wavelength of the generated signal and idler, and the temperature of the WGMR. The calculations are performed for a \SI{5.8}{\percent} MgO-doped lithium niobate WGMR and a wavelength range of the parametric light where high quality factors $\textrm{Q}>10^7$ can be expected (approximately \SIrange{500}{2500}{\nm}) \cite{Leidinger2015a}. Validity of the Sellmeier equation was tested over a temperature range from \SI{25}{\degreeCelsius} to \SI{250}{\degreeCelsius} in Ref.~{\cite{Schlarb1994}}. Smaller resonators exhibit stronger waveguide dispersion according to the increasing weight of higher-order terms in the dispersion (see Eq.~\ref{eq:disprel}). Degenerate signal and idler are already found at room temperature for a WGMR radius of $R=\SI{0.44}{\mm}$ at a pump wavelength of \SI{532}{\nm} (see Fig.~\ref{fig:tempsizes}(a)) and for a WGMR radius of $R=\SI{2.5}{\mm}$ at a pump wavelength of \SI{521}{\nm} (see Fig.~\ref{fig:tempsizes}(b)). 

The wide tuning range of the generated light from the visible to the near infrared allows for addressing a variety of physical systems indicated in Fig.~\ref{fig:tempsizes}. Especially relevant are those that have natural linewidths at the MHz scale which allows for highly efficient photon coupling. A swapping of the bipartite quantum state of the photon pair to two other physical systems, e.g. an optomechanical cavity \cite{Aspelmeyer2014} in a telecom band and a single atom, is possible for specific combinations of WGMR radius and pump wavelength.

\FloatBarrier
\subsection{Continuous tuning of the parametric frequencies}
\label{sec:continuous}
In this section we supplement the discussion of the phase matching temperature dependence by a second  independent mechanism, which may be realized via the linear electro-optical effect in lithium niobate \cite{Guarino2007,Michael2013}, by applying external pressure \cite{Ilchenko1998,Wagner2013} or by perturbing the resonator's evanescent field \cite{Teraoka2006,Schunk2015a}. This is required to continuously tune the generated signal or idler to an arbitrary frequency, e.g. an atomic transition, with MHz precision. Our concept for continuous tuning starts at a temperature phase matching point with zero PDC frequency mismatch $\Delta$ given by Eq.~\ref{eq:SPDCfrequ}. In a second step the temperature of the WGMR is changed, which initially leads to an increased PDC frequency mismatch. This acquired PDC frequency mismatch is then compensated with the second tuning mechanism resulting in phase matching at different frequencies of the modes. In a final step the pump laser frequency is tuned to the new resonance frequency of the same pump mode. In the following, we explain phase matching in case of the two tuning mechanisms and discuss various experimental implementations for the second tuning mechanism.

Continuous tuning of the parametric frequencies with temperature alone is limited to a narrow frequency interval determined by the differential thermal dispersion and bandwidths of the pump, signal, and idler modes. In our case, this frequency interval is of the same order as the bandwidths of the signal and the idler. For a type-0 quasi-phase matched PDC process in lithium niobate, where all three fields are extraordinarily polarized, tuning ranges can exceed the parametric bandwidths by an order of magnitude \cite{Werner2015}, because the resonance frequencies of pump, signal, and idler have a similar temperature dependence. The main downside of this method is the change in conversion efficiency (see Eq.~\ref{eq:conveffabove}) due to the temperature-dependent change in the PDC frequency mismatch $\Delta$. Continuous tuning without a change in conversion efficiency can be achieved by amending temperature tuning with a second tuning mechanism independent from temperature.

\begin{figure}[htb]
	\centering
	\includegraphics[scale=1]{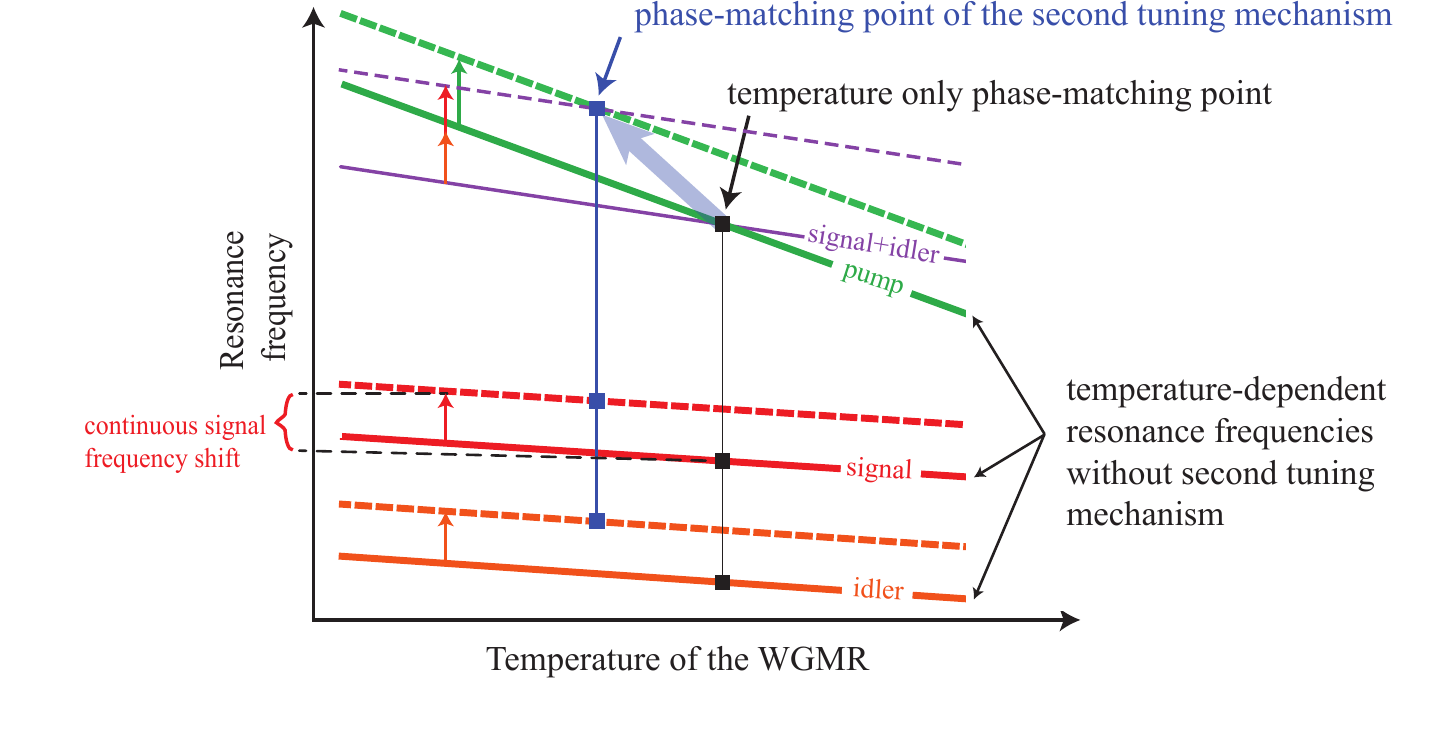}
	\caption{Continuous tuning of phase matching in a triply-resonant WGMR based on two independent tuning mechanisms. We assume that the examined mode triplet obeys the conditions of momentum conservation given by Eq.~\ref{eq:phasematch}. The temperature-dependent resonance frequencies $\nu(\ell_\textrm{p,s,i},\textrm{q}_\textrm{p,s,i},\textrm{p}_\textrm{p,s,i})$ of the pump, the signal, and the idler modes are shown as bold green, bold red, and bold orange lines, respectively. Additionally, the sum of signal and idler resonance frequencies $\nu(\ell_\textrm{s},\textrm{q}_\textrm{s},\textrm{p}_\textrm{s}) + \nu(\ell_\textrm{i},\textrm{q}_\textrm{i},\textrm{p}_\textrm{i})$ is shown as a thin purple line to illustrate the condition of energy conservation (see Eq.~\ref{eq:phasematcha}). Energy conservation for a mode triplet is fulfilled at any intersection of $\nu(\ell_\textrm{p},\textrm{q}_\textrm{p},\textrm{p}_\textrm{p})$ and $\nu(\ell_\textrm{s},\textrm{q}_\textrm{s},\textrm{p}_\textrm{s}) + \nu(\ell_\textrm{i},\textrm{q}_\textrm{i},\textrm{p}_\textrm{i})$ (see Fig.~\ref{fig:tempillstration}). In this example, the second tuning mechanism shifts the resonance frequencies in vertical direction as indicated by the arrows. This brings the intersection point to a different temperature and frequency pair. }
\label{fig:conttuning}
\end{figure}

Phase matching in case of two tuning methods is illustrated in Fig.~\ref{fig:conttuning}. The temperature-dependent resonance frequencies $\nu(\ell_\textrm{p,s,i},\textrm{q}_\textrm{p,s,i},\textrm{p}_\textrm{p,s,i})$ of one triplet of pump, signal, and idler modes are illustrated as bold lines. These modes obey the conditions of momentum conservation given by Eq.~\ref{eq:phasematch}. Energy conservation in the case of zero detuning $\delta_\textrm{p,s,i}$ is fulfilled at any intersection of the pump frequency $\nu(\ell_\textrm{p},\textrm{q}_\textrm{p},\textrm{p}_\textrm{p})$ with the sum of the signal and the idler frequency $\nu(\ell_\textrm{s},\textrm{q}_\textrm{s},\textrm{p}_\textrm{s}) + \nu(\ell_\textrm{i},\textrm{q}_\textrm{i},\textrm{p}_\textrm{i})$ (thin purple line), which occurs at exactly one point when only temperature is used for tuning. The second tuning mechanism shifts the resonance frequencies vertically leading to a new phase matching point at a different temperature and different resonance frequencies. In Fig.~\ref{fig:conttuning}, the pump and the parametric modes are blue shifted. In general other combinations of shifts are possible.

\begin{figure}[htb]
	\centering
	\includegraphics[scale=1]{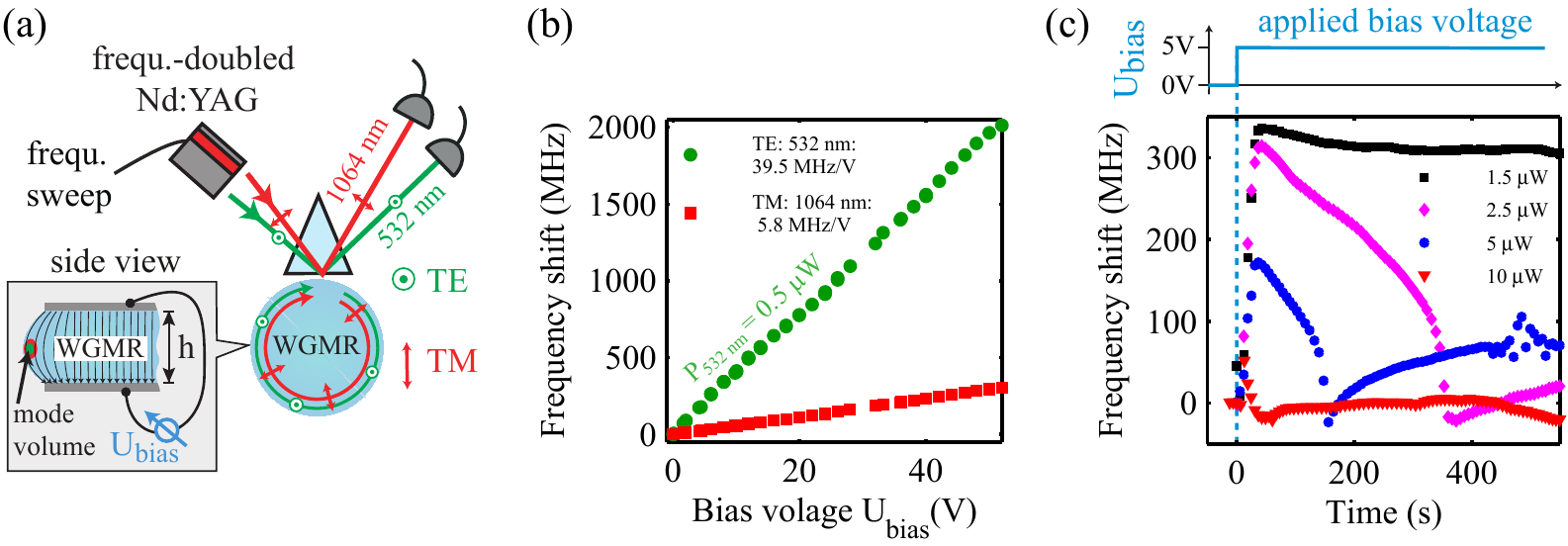}
	\caption{Tuning of the WGMR resonance frequencies with the linear electro-optic effect. (a) The top and bottom sides of the WGMR are coated with silver to apply a bias voltage $\textrm{U}_\textrm{bias}$. The resonator was excited in TE polarization to study the pump at a wavelength of \SI{532}{\nm} and in TM polarization at a wavelength of \SI{1064}{\nm} to study the parametric beams. (b) Nearly linear frequency shifts are obtained at low powers ($P_\textrm{532\,nm}=\SI{0.5}{\micro\watt}$) of the green pump and also for the infrared probe. (c) Demonstration of the frequency shift degrading at higher green pump power. At time $t=\SI{0}{\s}$ we apply a constant voltage $U_\textrm{bias}=\SI{5}{\volt}$ to measure the time-dependence of the linear electro-optic effect at different pump powers. A relaxation of the frequency shifts at the seconds scale is observed for green pump powers exceeding a few \SI{}{\micro\watt}. We attribute this effect to free charges induced via photoconductivity \cite{Gerson1986,Buse1999,Villarroel2010}, which lead to a local compensation of the external bias voltage around the mode volume.}
\label{fig:biasvoltage}
\end{figure}

The linear electro-optical effect in lithium niobate \cite{Weis1985} has been demonstrated to provide one possibility to tune the resonance frequencies of a monolithic resonator \cite{Guarino2007,Michael2013}. The refractive index change for signal and idler can be approximated as $\Delta n_\textrm{p,s,i} =  - \frac{1}{2} \, r_\textrm{p,s,i} \, n_\textrm{p,s,i}^3 \, {U_\textrm{bias}}/{h}$ with a voltage $U_\textrm{bias}$ applied over a distance $h$. The electro-optic coefficients in lithium niobate for the pump and the parametric modes are given by $r_\textrm{p} = \SI{31}{\pm\per\volt}$ and $r_\textrm{s,i} = \SI{8}{\pm\per\volt}$ for a voltage applied along the z-direction of the crystal \cite{Zgonik2002}.

In the experiment illustrated in Fig.~\ref{fig:biasvoltage}(a), we study the possibility of electro-optical tuning of PDC by measuring the voltage-induced frequency shifts of the modes with an external laser for the pump (TE, extraordinarily polarized) mode at a wavelength of \SI{532}{\nm} and for the signal and idler mode (TM, ordinarily polarized) at a wavelength of \SI{1064}{\nm}. We use a resonator made of MgO-doped lithium niobate (\SI{6.2}{\percent}) of radius $R = \SI{1.5}{mm}$, rim curvature $\rho = \SI{0.4}{mm}$, and thickness $h=\SI{0.5}{\mm}$ in this measurement. The top and bottom sides of the resonator are coated with silver. We expect electro-optical tuning rates of \SI{89}{\MHz\per\volt} and \SI{14}{\MHz\per\volt} for the pump and the degenerate parametric mode, respectively. The experiment was carried out at room temperature and at the pump power of $P_\textrm{532 nm}=\SI{0.5}{\micro\watt}$. The measured electro-optical tuning rates for the pump mode and the parametric mode of \SI{39.5}{\MHz\per\volt} and \SI{5.8}{\MHz\per\volt}, respectively (see Fig.~\ref{fig:biasvoltage}(b)), are reduced by approximately a factor of two compared to the theoretical values. One possible explanation is the fringe field effect, as illustrated in inset of Fig.~\ref{fig:biasvoltage}(a). A broadening of the bandwidths due to the conductive coating was not observed in the investigated voltage regime.

It should be pointed out that the electro-optical tuning rates in lithium niobate have the same sign for the TE and TM modes, and their ratio is close to the temperature tuning rates ratio. This limits the utility of the electro-optical tuning as the second mechanism. Moreover, photoconductivity and the onset of a photovoltaic current impedes the linear electro-optical effect \cite{Gerson1986,Buse1999} at increased green pump powers. Free charges are created within the mode volume leading to a local compensation of the electric field. A similar behavior for the TM-polarized beam at \SI{1064}{\nm} was not observed. For a green pump power of $P_\textrm{532\,nm}=\SI{10}{\micro\watt}$, the effect of an applied voltage already decays at the seconds scale (see Fig.~\ref{fig:biasvoltage}(c)). The bandwidth at critical coupling was approximately $\gamma_\textrm{532\,nm}=\SI{28}{\MHz}$ with \SI{70}{\percent} of transmitted pump power at the cavity resonance. A use of the electro-optical effect at a wavelenth of $\SI{532}{\nm}$ is therefore only possible in case of low pump powers or fast switching voltages.

Future studies may also involve frequency tuning by mechanical pressure, which causes a change in both the refractive index and the geometry of the WGMR. For microsphere resonators \cite{Ilchenko1998}, this technique has been demonstrated to provide frequency shifts over more than a free spectral range. In piezoelectric lithium niobate, however, the stress leads to electric fields, and the aforementioned electro-optic tuning becomes an inherent part of the stress tuning. A rigorous analysis of these phenomena combined effect has yet to be done.

For our tuning to atomic transitions, we chose another tuning mechanism involving a movable dielectric substrate placed within the evanescent field of the resonator. With this we demonstrated a tuning of the signal frequency over approximately \SI{400}{\MHz} (see Fig.~3(b) in \cite{Schunk2015a}). Frequency shifts induced by a dielectric placed within the evanescent field of the resonator \cite{Teraoka2006} are well-known from sensing experiments with WGMRs \cite{Vollmer2008,Sedlmeier2014}. A reduction of the optical quality factor via direct out-coupling of the intracavity light is avoided since the refractive index of the WGMR is higher than the refractive index of the dielectric substrate ($\textrm{n}_\textrm{ZnO}=2.03$ at \SI{532}{\nm} \cite{Bond1965}). The exact frequency shifts depend on the geometry and different refractive indices of the WGMR and the substrate, which is an ongoing field of research. The demonstrated tuning range, however, is sufficient to continuously bridge the frequency steps of the temperature fine tuning (see Fig.~\ref{fig:tempillstration}). This makes our WGMR-based source of narrowband photon pairs readily available for precision spectroscopy of MHz-wide transitions of arbitrary frequencies within the low-absorption window of lithium niobate \cite{Leidinger2015a} from \SI{500}{\nm} to \SI{2500}{\nm}.

\FloatBarrier
\section{Applications to narrowband atomic systems}
\subsection{Atomic spectroscopy above the OPO threshold}
In this section we show coupling of the signal light above the OPO threshold to two alkali species. We use the atoms as a reference to measure the absolute frequency and long-term stability of the signal frequency.

\begin{figure}[htb]
	\centering
	\includegraphics[scale=1]{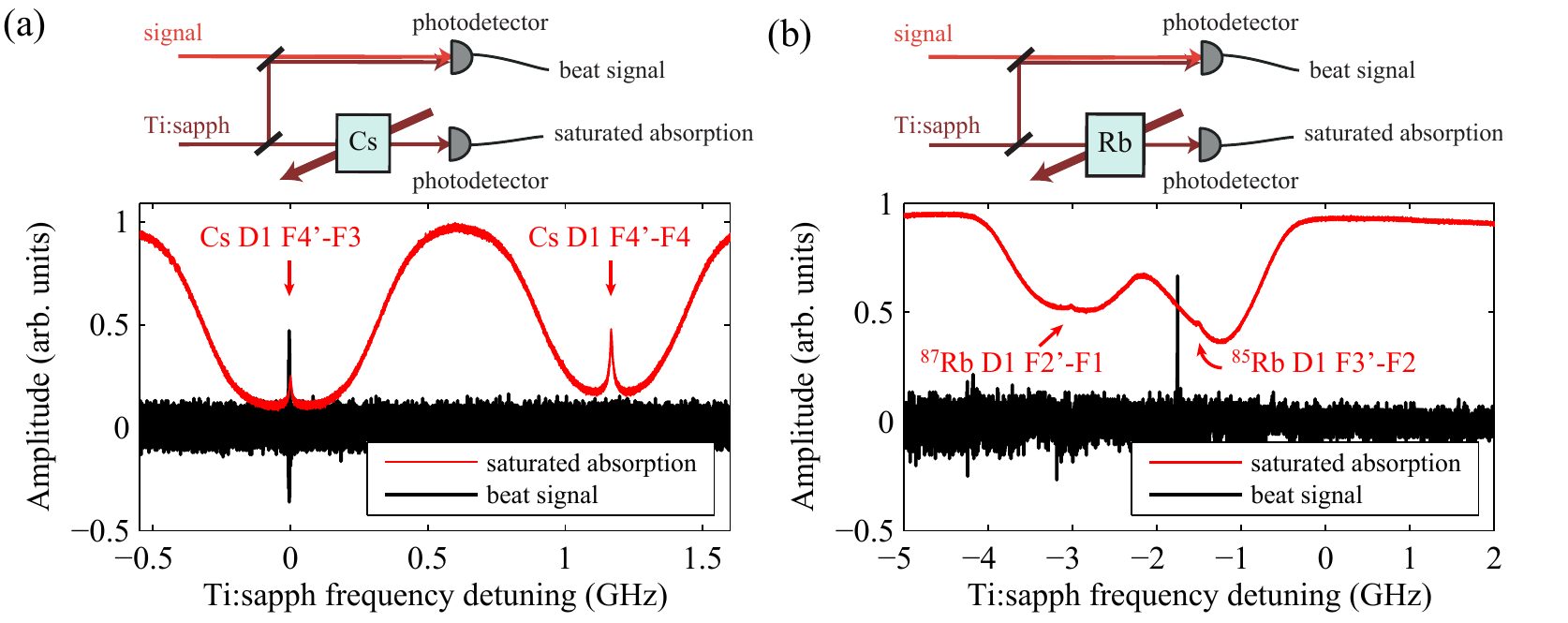}
	\caption{Atomic spectroscopy with PDC above threshold. The signal from PDC is on resonance with the D1 lines of cesium (a) and rubidium (b) (see Sec.~\ref{sec:selmodetrip} and \ref{sec:continuous} for the frequency tuning methods). We use a scanning Ti:sapphire reference laser in a saturated absorption configuration as a frequency reference at the two alkaline transitions. The peaks from saturated absorption indicate the respective transition frequencies in absence of Doppler-broadening.}
\label{fig:beatsignals}
\end{figure}

To generate PDC at the D1 lines of cesium and rubidium \cite{Schunk2015a}, we first selected a nearly resonant PDC mode triplet by temperature (see Sec.~\ref{sec:selmodetrip}) and then tuned the signal frequency continuously (see Sec.~\ref{sec:continuous}) to exactly the atomic resonance. The frequency of the generated signal can be measured at the MHz scale by beating it with a reference laser (SolsTiS cw Ti:sapphire laser from M Squared Lasers LTD). The frequency scan of the reference laser over the Doppler-broadened D1 lines of cesium in Fig.~\ref{fig:beatsignals}(a) and rubidium in Fig.~\ref{fig:beatsignals}(b) is used for an absolute frequency calibration. The cells containing the atomic vapors were \SI{5}{\cm} long and kept at a temperature of \SI{80}{\degreeCelsius}. The employed method of sweeping the reference laser provides a large dynamic range of approximately \SI{20}{\GHz} for the measured frequency.

\begin{figure}[htb]
	\centering
	\includegraphics[scale=1]{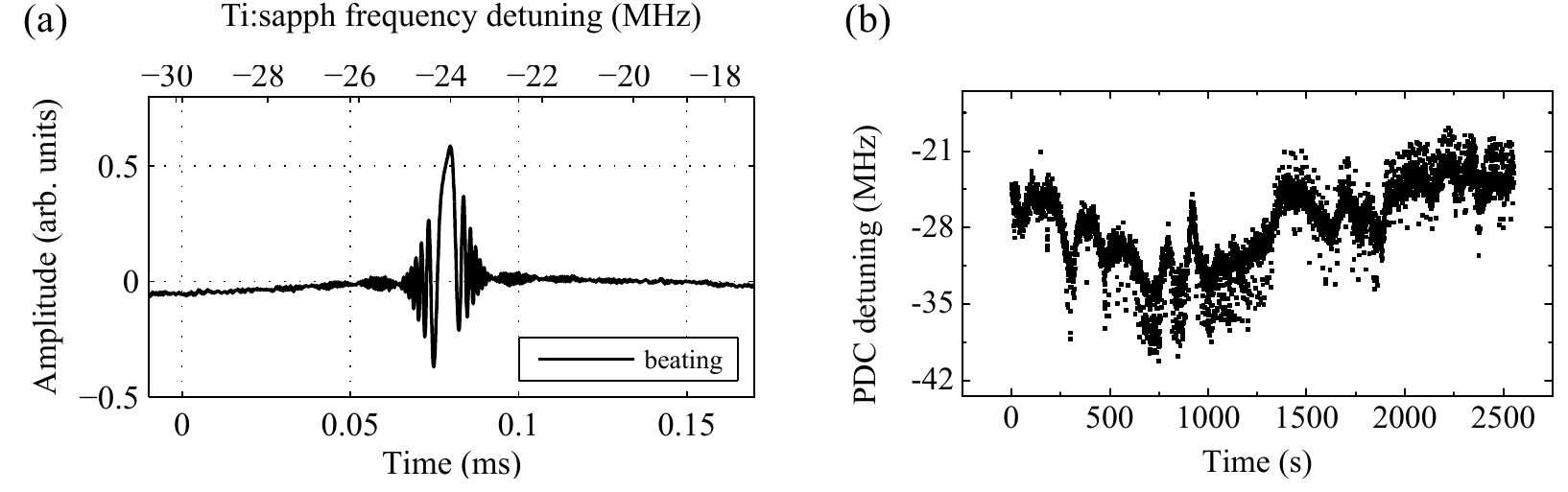}
	\caption{(a) A beat node between the generated signal and the reference laser is used to measure the signal central frequency at MHz-accuracy relative to the Cs D1 F4'-F3 transition. In contrast to the measurement shown in \ref{fig:beatsignals}(a), we used a \SI{1}{\MHz} low-pass filter for noise suppression. (b) The fluctuations of the signal frequency are observed during \SI{43}{min.} with approximately two measurements per second. This demonstrates a frequency stability at the MHz scale required for efficient photon-atom coupling.}
\label{fig:stabiilty}
\end{figure}

In Fig.~\ref{fig:stabiilty}(a) a zoom of the measured beat signal at the Cs D1 line (see Fig.~\ref{fig:beatsignals}(a)) is shown. The frequency sweep of the reference laser provides the time axis below and the frequency axis on top. The beat note is detected if it has frequency components below the bandwidth of the detector (\SI{11}{\kHz}). This requires that the frequency difference between the reference laser and the signal light central frequencies do not exceed the larger of their bandwidths. The reference laser has a bandwidth of \SI{50}{\kHz}. Hence, the frequency width of the beat note gives an upper bound for the bandwidth of the generated signal, which is well below one MHz. From theory, this bandwidth is determined by the pump laser bandwidth \cite{Debuisschert1993,Kozyreff2008}, which is approximately \SI{1}{\kHz} in the above threshold case. In contrast, the bandwidth of the signal mode is several MHz, which we investigate below threshold via signal-idler cross-correlation measurements (see Fig.~\ref{fig:BWtuning}).

The beat note shown in Fig.~\ref{fig:stabiilty}(a) is recorded for a longer period to demonstrate the frequency stability of our source. A short-time stability of about \SI{3.5}{\MHz} (\SI{10}{\second} integration time) is achieved. This frequency stability is already below the natural linewidth $\Gamma \approx \SI{4.58}{\MHz}$ of the Cs D1 line \cite{Steck2008all}. The temperature of the WGMR is measured with a thermistor and stabilized with a proportional-integral-derivative controller with millikelvin precision. More advanced temperature locking techniques based on simultaneous measurements of two eigenmodes of the WGMR can allow for a temperature control with nanokelvin precision based on a simultaneous measurement of two orthogonal eigenmodes of the WGMR \cite{Strekalov2011,Weng2014}.

\FloatBarrier 
\subsection{Atomic spectroscopy below the OPO threshold}
In this section we demonstrate one of the most basic applications of heralded photons matched in bandwidth and central wavelength to an atomic transition. Here we switch from linear photodetectors to Geiger-mode detection using avalanche photo diodes (APDs). For the idler photons we use ID220 from ID Quantique (APD 1) and for the signal photons SPCM CD 3017 from Perkin Elmer (APD 2). A stochastic absorption-emission of a heralded signal photon by an optical transition of an atom modifies its temporal correlation function with the heralding idler photon in a way that combines the optical transition and the resonator ring down times. This experiment was initially reported in Ref.~\cite{Schunk2015a}. We believe this experiment demonstrates an interesting approach to shape single-photon pulses for quantum optics applications \citep{Maiwald2012,Bader2013} and for time-resolved single-photon spectroscopy of atomic transitions.

\begin{figure}[htb]
\centerline{
	\centering
	\includegraphics[scale=1]{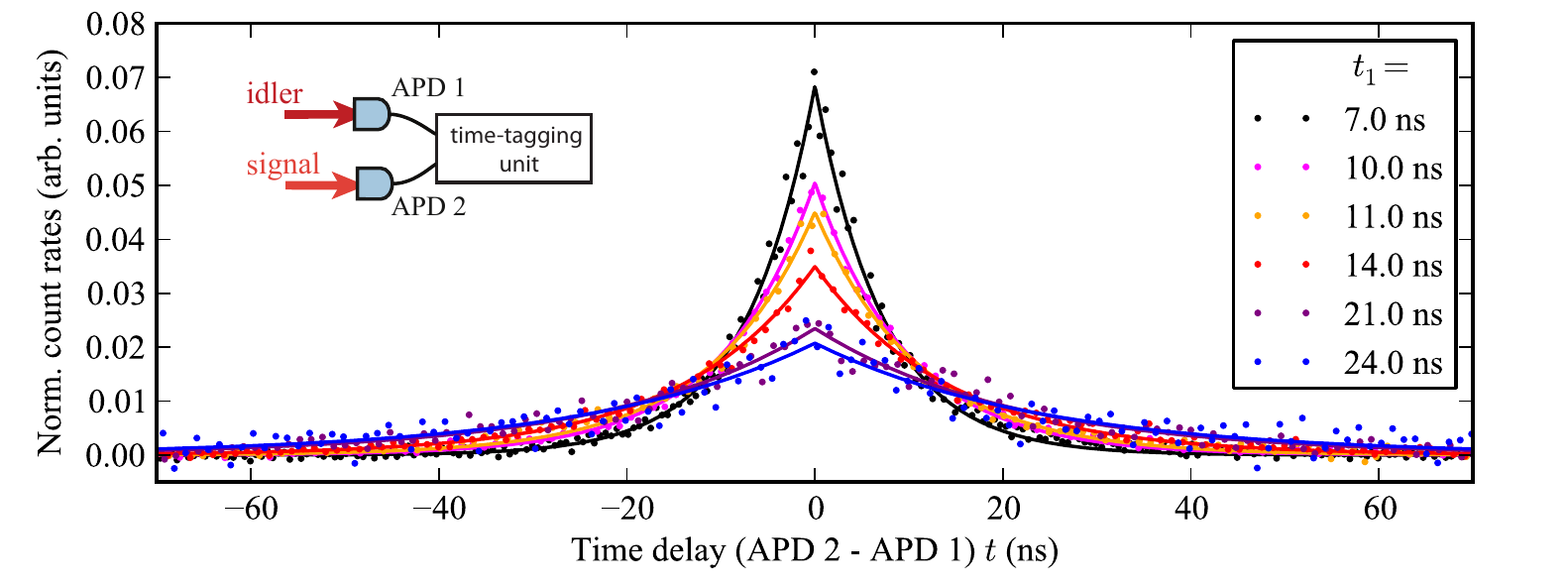} }
\caption{Bandwidth tuning of parametric down conversion. The coupling rate to the WGMR can be adjusted via the distance between the WGMR and the coupling prism, which directly affects the bandwidth of the photon pairs. A double exponential fit $\propto \textrm{exp}(-\vert t \vert / t_1)$ \cite{Michael2013} of the normalized count rates was used to determine the ring down time $t_1$ and bandwidths $\gamma_\textrm{si} = 1 / (2 \pi \cdot t_1)$.}
\label{fig:BWtuning}
\end{figure}

The signal-idler temporal correlation function is initially determined by the resonator ring down times  $t_s$ and $t_i$ for the signal and idler modes, respectively. If these times are equal, $t_i=t_s \equiv t_1$, the correlation function takes the following form \cite{Ou1999,Michael2013,Luo2015}:
\begin{align}
	f(t)=\frac{1}{2t_1} \exp{\left(-\frac{|t|}{t_1}\right)} \,.
	\label{corf_ini}
\end{align}
This function is normalized for unity probability:
\begin{align}
	\int_{-\infty}^{\infty}  f(t) \, \mathrm{d}t=1 \,.
	\label{eq:norm}
\end{align}
In WGMRs, the ring down time $t_1$ can be continuously tuned by changing the output coupling rate \cite{Michael2013}. In Fig.~\ref{fig:BWtuning} we show several examples of the signal-idler correlation functions observed in our experimental setup under different coupling conditions. Each curve is normalized according to Eq.~\ref{eq:norm}.

\begin{figure}[htb]
\centerline{
	\centering
	\includegraphics[scale=1]{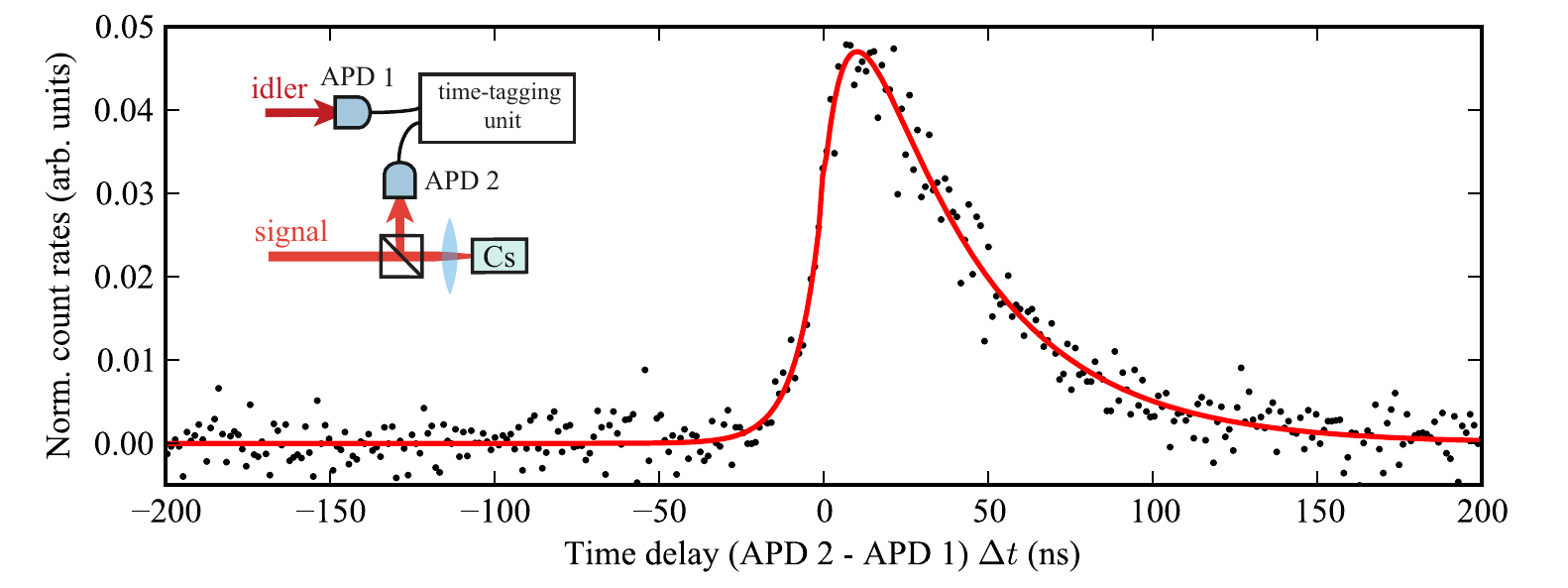} }
	\caption{Time-correlation function for a heralded photon emitted by cesium D1 transition. Data (black dots) collected for one hour with 1 ns resolution is consistent with the theoretical curve (red line) given by Eq.~\ref{eq:marg} derived on the basis of classical probability analysis.}
\label{fig:g2}
\end{figure}

The signal-idler correlation function modified by a single absorption-emission cycle of the signal photon is shown in Fig.~\ref{fig:g2}. To ensure that the scattering took place we collect photons in backward direction as shown in the inset of Fig.~\ref{fig:g2}. To suppress multiple scattering events, the measurement was performed at low optical density when no radiation trapping occurred. Note that the opposite limit of high optical density may lead to the heralded single photon super-radiance \cite{Oliveira2015}.

The effect of a single absorption-emission cycle on the signal-idler correlation (Eq.~\ref{corf_ini}) is adequately described in terms of a classical probability distribution. After re-emission, the signal has an additional delay $\tau$ with the following distribution function:
\begin{align}
	g(\tau) =\frac{1}{t_2}
		\begin{cases}
			0 & \quad{\tau<0}   \\
			\exp{(-\frac{\tau}{t_2})} & \quad{\tau\geq 0} \,,
		\end{cases}
\end{align}
where $t_2$ is the atomic excited state life time. The probability density $g(\tau)$ is also normalized:
 \begin{align}
	\int_{-\infty}^{\infty}  g(\tau) \, \mathrm{d}\tau=1 \,.
 \end{align}
The resonator and atomic decay events are statistically independent, so the composite distribution function of delay times $t$ and $\tau$ is given as a		product:
 \begin{align}
	F(t,\tau)=f(t) \cdot g(\tau) \,.
 \end{align}
Let us now introduce symmetric time variables $t'$ and $T$ so that $t=(t'+T)/\sqrt{2}$ and $\tau=(t'-T)/\sqrt{2}$.
Note that $\Delta t = \sqrt{2}t'=t+\tau$ is the time delay between the APD2 and APD1 firing, which is measured in the experiment.

We obtain
\begin{align}
	F(t',T) =\frac{1}{2t_1t_2}\exp{\left(-\frac{|t'+T|}{\sqrt{2}t_1}\right)} 
		\begin{cases}
			0, & \quad{T>t'}   \\
			\exp{(-\frac{t'-T}{\sqrt{2}t_2})}, & \quad{T\leq t'} \,,
		\end{cases}
\end{align}
and
\begin{align}
	\int_{-\infty}^{\infty}\int_{-\infty}^{\infty}  F(t',T) \, \mathrm{d}T\mathrm{d}t' = 1.
\end{align}
From $F(t',T)$ we find the marginal probability distribution $P(t')$:
\begin{align}
	P(t') &=\int_{-\infty}^{\infty}  F(t',T) \, \mathrm{d}T \nonumber 	\\ &=  \frac{1}{2t_1t_2}\int_{-\infty}^{t'} \exp{\left(-\frac{|t'+T|}	{\sqrt{2}t_1}\right)}\exp{\left(\frac{T-t'}{\sqrt{2}t_2}\right)}	\mathrm{d}T.
	\label{eq:marg}
\end{align}
Computing the integral in Eq.~\ref{eq:marg} and substituting  $P(t')=P(\Delta t/\sqrt{2})/\sqrt{2}$ (dividing the probability density by $\sqrt{2}$ preserves its normalization) in the result, we arrive at
\begin{align}
	P(\Delta t) =\frac{1}{2} 
		\begin{cases}
			\frac{\exp{(\frac{\Delta t}{t_1})}}{t_1+t_2}  & \Delta t<0 \\
			\frac{\exp{(-\frac{\Delta t}{t_2}})}{t_1+t_2}+\frac{\exp{(-\frac{\Delta t}{t_1})}-\exp{(-\frac{\Delta t}{t_2}})}{t_1-t_2} & 	\Delta t\geq 0 \,.
		\end{cases}
		\label{new_prob}
\end{align}
Note that despite its segmented structure, $P(\Delta t)$ remains analytic at $\Delta t = 0$.

We use Eq.~\ref{new_prob} to fit of the correlation data in Fig.~\ref{fig:g2}. The least-square fitting yields the resonator ring-down time $t_1 = 7.4$ ns (21.5 MHz bandwidth) and the atomic excited state decay time $t_2 = 37$ ns (4.3 MHz bandwidth) consistent with the natural linewidth $\SI{4.58}{\MHz}$ of the Cs D1 line. We measured a Klyshko efficiency \cite{Klyshko1980} of \SI{8.4}{\percent} for the idler photons and \SI{0.067}{\percent} for the re-emitted signal photons from fluorescence. The Klyshko efficiencies in our experiment are currently limited by the ratio of coupling rate and bandwidth of the single photons, the fiber coupling efficiency, the relatively small collection angle (NA=0.4) for the photons from fluorescence, and the detection efficiency of the APDs.

\section{Conclusion}
In this work we have demonstrated the addressing of different atomic transitions with a narrowband photon pair source based on cavity-assisted parametric down-conversion. To achieve this, we characterized the central frequencies and bandwidths of the parametric light generated by our source, which is a triply-resonant whispering-gallery mode resonator made of lithium niobate. This source allows for stepwise tuning of the parametric frequencies over hundreds of nanometer by selecting specific mode triplets by temperature, and continuous tuning by amending temperature tuning with a second independent tuning mechanism. Resonators are manufactured from lithium niobate wafers by single point diamond turning and polishing techniques leading to quality factors of $10^{7}-10^{9}$. A future implementation of this source in large-scale quantum repeater networks will require on-chip manufacturing, which currently reach quality factors of $10^{5}$ for lithium niobate \cite{Guarino2007}. Further studies may involve different resonator materials, different phase matching configurations such as cyclic phase matching \cite{Lin2013}, a coating of the WGMR to achieve a radially inhomogeneous refractive index \cite{Chowdhury1991} or advances in periodic poling of WGMRs \cite{Beckmann2011,Meisenheimer2015}, which is already a standard technique for linear crystals. Future studies may also improve the experimental control over the temporal and frequency profile by pulsed excitation to further optimize the applicability of WGMR based sources of quantum light.

\section*{Acknowledgments}
The authors kindly acknowledge the support from Michael F\"{o}rtsch, Navid Soltani, Ralf Keding, Andrea Cavanna, and Felix Just.

\section*{Funding information}
The authors are grateful for the financial support of the European Research Council under the Advanced Grant PACART. Ch. M. acknowledges support from the Alexander von Humboldt Foundation.

%
%

\end{document}